\documentclass[12pt, preprint]{aastex63}

\usepackage{natbib}
\usepackage{booktabs}
\usepackage{rotating, graphicx}

\def\lax {\ifmmode{_<\atop^{\sim}}\else{${_<\atop^{\sim}}$}\fi}
\def\gax {\ifmmode{_>\atop^{\sim}}\else{${_>\atop^{\sim}}$}\fi}
\def\gtorder{\mathrel{\raise.3ex\hbox{$>$}\mkern-14mu
             \lower0.6ex\hbox{$\sim$}}}

\def\s1{s$^{-1}$}
\def\cm2{cm$^{-2}$}
\def\ster1{ster$^{-1}$}

\shorttitle{Comptonization spectra of nmCVs}
\shortauthors{Maiolino et al.}

\begin{document}

\title{Testing Comptonization as the origin of the continuum 
in nonmagnetic Cataclysmic Variables. The photon index of X-ray emission}

\author[0000-0002-4918-7182]{T. Maiolino}
\affiliation{School of Physics and Technology, Wuhan University, Wuhan
430072, China}
\affiliation{WHU-NAOC Joint Center for Astronomy, Wuhan University, 
Wuhan 430072, China}
\affiliation{Dipartimento di Fisica, Universit\`a di Ferrara, via Saragat
1, 44122 Ferrara, Italy}

\author{L. Titarchuk}
\affiliation{Dipartimento di Fisica, Universit\`a di Ferrara, via Saragat
1, 44122 Ferrara, Italy}
\affiliation{Astro Space Center, Lebedev Physical Institute, Russian
Academy of Sciences,  Profsouznay ul. 84/32, Moscow 117997, Russia}

\author[0000-0003-1304-9914]{F. D'Amico}
\affiliation{Instituto Nacional de Pesquisas Espaciais, 
Avenida dos Astronautas 178, 12227-010 S.J. dos Campos-SP, Brazil}

\author[0000-0002-9983-8609]{Z. Q. Cheng}
\affiliation{School of Physics and Technology, Wuhan University, Wuhan
430072, China}
\affiliation{WHU-NAOC Joint Center for Astronomy, Wuhan University,
Wuhan 430072, China}

\author[0000-0003-3901-8403]{W. Wang}
\affiliation{School of Physics and Technology, Wuhan University, Wuhan
430072, China}
\affiliation{WHU-NAOC Joint Center for Astronomy,
Wuhan University, Wuhan 430072, China}

\author[0000-0003-0946-3151]{M. Orlandini}
\affiliation{INAF/OAS Bologna, via Gobetti 101, 40129 Bologna, Italy}

\author[0000-0003-2284-571X]{Filippo Frontera}
\affiliation{Dipartimento di Fisica, Universit\`a di Ferrara, via Saragat
1, 44122 Ferrara, Italy}
\affiliation{INAF/OAS Bologna, via Gobetti 101, 40129 Bologna, Italy}
\affiliation{ICRANET Piazzale d. Repubblica 10-12, 65122 Pescara (PE), Italy}

\correspondingauthor{Tais Maiolino}
\email{tais.maiolino@whu.edu.cn}
   
\begin{abstract}
X-ray  spectra of nonmagnetic cataclysmic variables (nmCVs) in the $\sim$ 0.3$-$15 keV energy band have been described either by one or several optically thin thermal plasma components, or by cooling flow models.
We tested if the spectral continuum in 
nmCVs could be successfully described by Comptonization of soft photons off hot electrons presented in a cloud surrounding the source [the transition layer, (TL)].
We used publicly \textit{XMM-Newton} Epic-pn, \textit{Chandra} HETG/ACIS and LETG/HRC, and \textit{RXTE} PCA and HEXTE observations of four Dwarf Novae (U~Gem, SS~Cyg, 
VW~Hyi and SS~Aur) observed in the quiescence and outburst states. In total, we analyzed  18 observations, 
including a simultaneous 0.4-150 keV \textit{Chandra}/\textit{RXTE} spectrum of 
SS~Cyg in quiescence. We fitted the spectral continuum with up to two thermal Comptonization components  (\textsc{compTT} or \textsc{compTB} models in XSPEC),
using only one thermal plasma temperature and one optical depth. In this framework the two seed photon components are coming presumably from the innermost and outer parts of the TL (or innermost part of the disk).  
We obtained  that the thermal Comptonization can successfully describe the spectral continuum of these nmCV  in 
the $\sim$ 0.4$-$150 keV energy band.  Moreover, we present the first principal radiative transfer model which explains 
the quasi-constancy of the spectral photon index observed around 1.8, which strongly supports the Comptonization framework in nmCVs.
\end{abstract} 


\section{Introduction}

The power source of X-rays in cataclysmic variables (CVs) is known to be due to the 
accretion of matter onto a compact object -- a white dwarf (WD). In non-magnetic CVs 
(nmCVs), as well as in all CVs containing accretion disks, the accretion disk is 
generally too cold (kT $\ll$ 1 keV) to emit X-rays \citep{Lewin2006}. 
The major source 
of X-rays have been identified with the transition  layer (TL) -- the region between the 
spiralling accretion disk and the surface of a more slowly rotating WD, wherein half 
of the gravitational energy is expected to be released. Physical characteristics of 
the TL and the system have been described to account for both the observed soft 
and hard X-ray spectral emission \citep{Pringle1979,Patterson1985a,Patterson1985b,Titarchuk2014}.   

In the standard TL framework the accretion rate determines the TL optical depth, which along with the electron temperature $T_e$ of the TL drive the X-ray spectral emission. 
\citet[][and references therein]{Patterson1985a} 
suggested that  at high accretion rate ($\dot{M} \gtrsim 10^{16}$ g~s$^{-1}$), the TL is expected to 
be optically thick and to radiate a blackbody component at $T\sim10^5$~K. 
 On the contrary, at low accretion 
rate ($\dot{M} < 10^{16}$ g~s$^{-1}$) -- i.e., in the quiescent state -- the TL is 
expected to have a lower density and to be optically thin. Because optically thin 
plasma does not cool efficiently, higher temperatures are expected in this state. 
The cooling of the TL in this case has been described by a thermal bremsstrahlung 
component, emitting hard X-rays at temperature $T\sim10^8$~K \citep{Patterson1985b}. 
Part (up to a half) of the hard X-rays emitted by the TL must be absorbed by a WD surface, and reradiated in soft X-rays with an effective temperature 
$T_{eff}~\lesssim~10^{5-6}$~K \citep[see, e.g.,][]{Patterson1985b,Williams1987,Mukai2017}.

Early spectral analysis of nmCVs showed that their spectral continuum were generally 
well described by only one thermal bremsstrahlung component, i.e., by a single plasma 
temperature in the 1-5 keV energy range, and of a few keV up to $\sim$ 10 keV in the case of dwarf novae \citep[and  references therein]{Lewin2006}. 

In the last decades the single temperature \textsc{mekal} (or its variation \textsc{vmekal}) model \citep[see  XSPEC;][]{Mewe1986,Liedahl1995} has been broadly used  to fit the CVs spectra. In this model the X-ray radiation is produced by a hot, optically-thin thermal 
plasma: the basis of the continuum radiation is free-free (bremsstrahlung), free-bound, 
and two-photon emission. The emission lines observed are included into this model and 
are modeled by: excitation from electron impact, radiative and dielectronic recombination 
and by inner-shell excitation and ionization. As the model is in the optically thin limit, 
photo-ionization or photo-excitation effects are not taken into  account. Similar to the 
MEKAL model, the APEC code\footnote{Calculated using the \textsc{atomdb} code,
more information can be found at {http://atomdb.org}.} 
(or its variations \textsc{vapec} and \textsc{vvapec}) is another thermal optically 
thin plasma model which has been commonly used to describe the spectra of nmCVs. 

Spectral analysis using data from X-ray observatories with better signal-to-noise and 
energy resolution (e.g., \textit{XMM-Newton} and \textit{SUZAKU}) showed that only one 
optically thin plasma temperature does not satisfactory fit the spectra of all sources \citep{Lewin2006} 
-- there are analyses in which two or even more plasma temperatures are needed to obtain  good  
spectral fits \citep[see, e.g.,][]{Pandel2003}. For that reason, it has been suggested that the cooling flow spectral model should represent a more physically correct description of the observed X-ray spectra.

The gas flow in cooling flow models is assumed to be composed by a range of temperatures which 
vary from the hot shock temperature $kT_{\rm{max}}$ (the maximal temperature to that the plasma 
is heated up; $kT_{\rm{max}}~\sim10-80$~keV) to the temperature  when the optically thin cooling material  settles on to the WD surface \citep[see][and references therein]{Mukai2003,byckling2010}. 
Cooling flow models are available in XSPEC, as:  \textsc{cemekl} or \textsc{cevmkl}, \textsc{mkcflow} 
or \textsc{vmcflow} \citep{Mushotzky1988ASIC..229...53M} -- these models were originally developed to 
describe the cooling flows in clusters of galaxies. It is now known that such models
do not successfully describe these sources \citep[as stated in][]{Mukai2003}. All of them 
basically  use an optically thin thermal plasma model for the individual temperature components.  

\citet{baskill2005} used \textbf{} the \textsc{cemekl}  cooling flow model to successfully fit $\sim$30 
DNe spectra observed with \textit{ASCA} (their source sample includes three out of four of the 
nmCVs analyzed in this paper: SS~Cyg, U~Gem and VW~Hyi). \citet{byckling2010} through 
\textit{XMM-Newton}, \textit{Suzaku} and \textit{ASCA} data, fitted the spectra of 12 DNe 
(their source sample includes SS~Cyg, U~Gem and SS~Aur) using either one temperature 
optically thin plasma model or a 
cooling flow model. All fits include spectral components  taking into account absorption 
created by the presence of material along the line of sight. A photoelectric absorption column 
is usually used (e.g., the \textsc{wabs} model in XSPEC was extensively used in the last years), 
with  additional partial covering absorption (e.g., \textsc{pcfabs} in XSPEC) in some 
cases -- to take into account for the presence of an intrinsic absorber located somewhere 
within the binary system \citep[see, e.g.,][]{byckling2010}.

\citet{Mukai2003}, using {\it Chandra} high energy transmission grating (HETG) data, stated that 
there are two types of CV X-ray spectra and suggested that the spectral differences might lie 
on the specific accretion rate (accretion rate per unit of area) of the systems. They showed 
that all three nmCVs analyzed (and an unusual Intermediate Polar (IP), EX Hya) had their spectra better described 
by cooling flow models, whereas  other sources analyzed (which were all of the IP type) had 
their spectra better described by the  photo-ionization model (PHOTOION model in XSPEC). 

\begin{figure}
\centering
\includegraphics[width=0.9\textwidth]{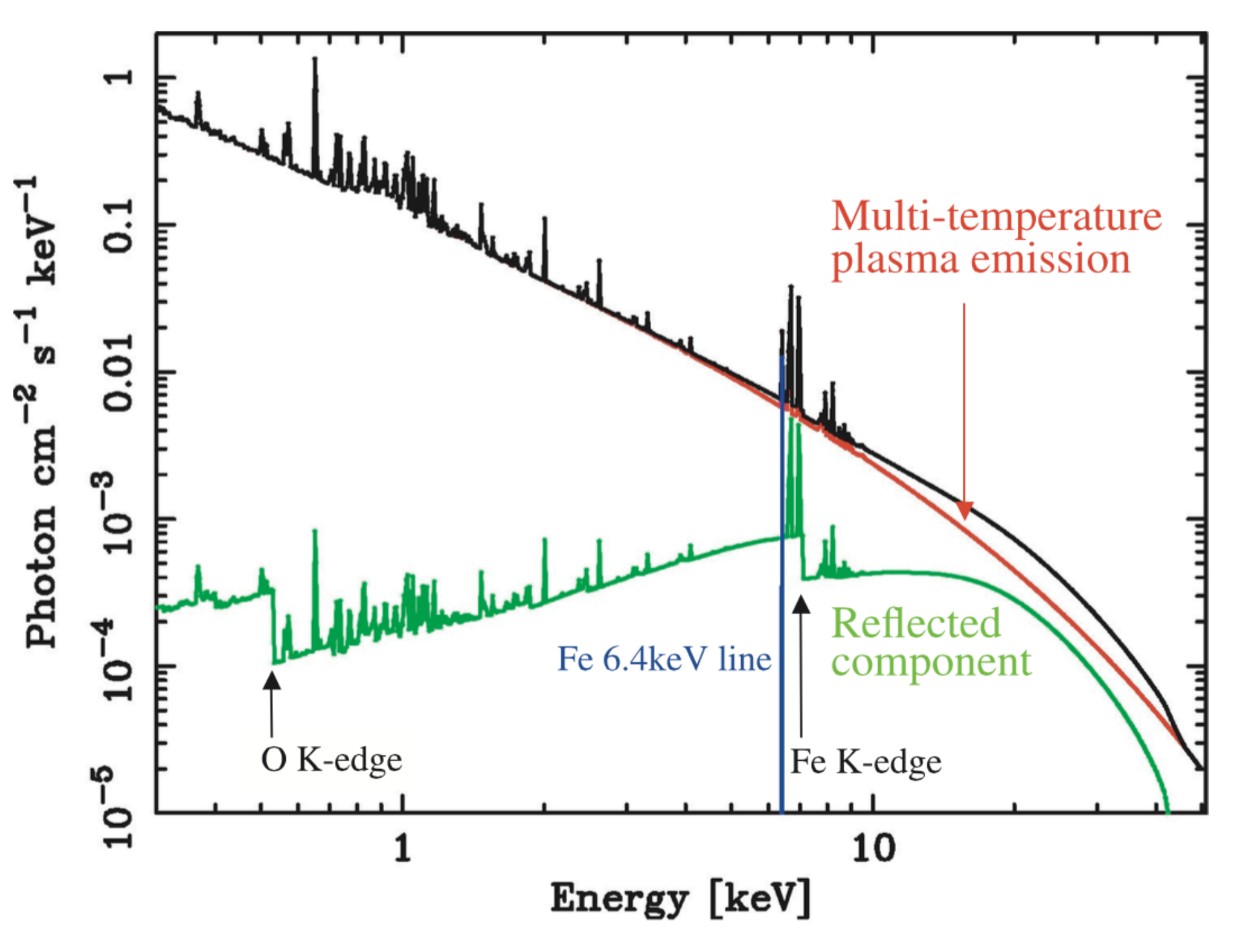}
\caption{The spectral components of SS~Cyg in 
the 0.2--40 keV band, which basically represents the spectral modeling of CVs: 
a multi-temperature optically thin thermal plasma represents the hard X-ray 
emission \textit{(red)}, the reflection of the hard X-rays by the accretion 
disc or/and the surface of the WD produces the fluorescent iron $K_\alpha$ 
lines observed and a hump in the continuum 
(green)
\citep{Done1997,ishida2009}.}
\label{fig:CVreflection}
\end{figure}

Figure \ref{fig:CVreflection} shows the spectral components of the nmCV SS~Cyg 
in the 0.2--40 keV energy band \citep{ishida2009}. This picture is proposed by \citet{Done1997} 
on the basis of their analysis of SS~Cyg using \textit{Ginga} and \textit{ASCA} data, and 
represents a general framework of the radiative process in CVs, widely 
accepted (by the time of this paper) by the scientific community. The red line in Figure 
\ref{fig:CVreflection} represents the hard X-ray spectrum -- primarily composed of an 
optically thin thermal plasma emission with a temperature distribution. In this framework, 
a part of the hard X-rays produced in the optically thin thermal plasma suffers reflection from 
the the accretion disk and/or the white dwarf surface, which represent a significant fraction 
of the observed X-ray flux. This reflected component (represented by  
\textit{green} line in Figure~\ref{fig:CVreflection}) explains, in the same way as in low mass X-ray binaries (LMXBs) 
and type I Seyfert galaxies  \citep{George1991}, the production of the neutral fluorescent iron $K_\alpha$ 
line observed at 6.4 keV in several CVs. The presence of this feature is explained by 
reflection of X-rays from the accretion disk and/or the WD surface, where relatively cold iron with 
temperatures $\leq 10^6 K$ is present. 
In general, up to three distinct iron emission lines in the $\sim$ 6.4 to 7.0 keV 
energy range may be present, in the spectra of CVs, related to neutral, 
He-like and H-like k$_\alpha$ Fe lines. The He-like and H-like lines (at $\sim$ 6.7 and 7.0 keV, respectively) 
are produced by a plasma with relatively higher temperatures of $10^{7-8}$ K \citep[see, e.g.][]{Rana2006b}.

For a more detailed review on nmCVs and their X-ray emission -- as well as on other types of CVs -- 
see for example \citet{Warner1995} and \citet{Mukai2017}.

{The Comptonization model is not one of the standard models which has been currently used to describe the CV continuum. However, CVs, mainly the non-magnetic and IP types, share structural similarities with LMXBs -- i.e., these systems contain: an accretion disk (which appears either entirely in nmCVs or partially in IPs), a transition  layer (TL, corona), and a compact object.} Taking into account these geometric/structural similarities and the radiative processes used to describe the continuum of LMXBs -- which has changed from bremsstrahlung to the Comptonization -- we tested if the spectral continuum of 
nmCVs and IPs could be successfully described by the Comptonization (up-scattering) of soft  photons off hot electrons of a Compton cloud around the compact object, as it is in LMXBs. In this paper 
we present the results of this verification performed on nmCVs. We used 4 \textit{XMM-Newton} Epic-pn, 8 \textit{Chandra} (LETG/HRC 
and HETG/ACIS), and 6 \textit{RXTE} (PCA and HEXTE) publicly observations of four Dwarf Novae (DNe): U~Gem, 
SS~Cyg, VW~Hyi and SS~Aur. These sources were observed only in quiescence by \textit{XMM-Newton}, and 
in the quiescence and outburst states by \textit{Chandra} and \textit{RXTE}. In our observational sample, 
we have a simultaneous \textit{Chandra}/\textit{RXTE} observation of SS Cyg, which provides us the 0.4$-$150 keV 
broadband spectral description of this source in the {quiescence} state. 

In a previous paper by \cite{Titarchuk2014}, the authors found that the photon index, $\Gamma$ in the NS sources strongly concentrated at 2. 
Moreover, they analytically demonstrated, using the radiative transfer model, that this is a result of the gravitational release of accreted material in the transition layer (TL) near  a NS surface.
Our goal is to show that the similar effect should be seen in  WDs in which we should take into account an absorption of the X-ray radiation by  the WD surface (while in a NS, its surface reflects X-ray radiation).  

This paper is organized as follows: in section \ref{sec:sources} we describe our source sample, 
in section \ref{sec:datared} we present the data reduction of the \textit{XMM-Newton} (Epic-pn), 
\textit{Chandra} (LETG/HRC and HETG/ACIS), and \textit{RXTE} (PCA and HEXTE) observations, 
in section \ref{sec:specanalysis} we describe the spectral analysis, in section 
\ref{sec:model}, 
we present our radiative transfer  model which explains the invariance  of the photon index,   in section  
\ref{sec:discussion} 
we discuss the results of the paper, and in section  \ref{sec:conclusions} 
we present our final conclusions.

\section{Source Sample} 
\label{sec:sources}

Since the goal of this work is to study the spectral continuum of nmCVs, 
any source belonging to this type  have been included in our sample. Because we have started this study 
using XMM-Newton Epic-pn data, we have chosen sources present in the archive with publicly Epic-pn observations.  
Thus  we included 4 nmCVs in our sample: U~Gem, SS~Cyg, VW~Hyi and SS~Aur, which are amongst the brightest X-ray nmCV.

For completeness, in this section we give a brief description of these sources. We present the binary system parameters [orbital period ($P_{\rm{orb}}$), rotational period ($P_{\rm{spin}}$), and an inclination angle ($i$)], distance, and temporal properties, and examples  of the spectral modeling published  in the literature for each source.  

\subsection{U Gem}
\label{sec:UGem}

U~Gem has a WD with mass ($M_{WD}$) of $1.07\pm0.08\ M_\odot$ \citep{Cat2003}. It appears to be at most slowly rotating 
($v\sin i\lesssim100$ km~s$^{-1}$), which corresponds to 20\% of the break-up velocity \citep{Lewin2006}. The secondary star has a mass $M_2$ of $0.39\pm0.02\ M_\odot$ \citep{Cat2003} and spectral type M4~V \citep{harrison2000}. The system is 
located at $100\pm4$ pc of distance; it has an orbital period of 4.246~h \citep{harrison2004} and an inclination 
$i$ of $69\pm2^{\circ}$ \citep{Cat2003}.  Dips in the X-ray light curves have been observed during quiescence \citep{Szkody1996,Szkody2002}, and during normal 
\citep{Long1996} and anomalously outbursts (of $\sim$45 days) \citep{Mason1988} as well.  
The X-ray dips in U~Gem are less deep in quiescence than in outburst. The morphology of the dips changes from cycle to 
cycle, related to changes in the absorbing material.
In U~Gem both soft and hard X-ray fluxes are higher during outburst than in quiescence by a factor of $\sim$10--100 
\citep[see][and references therein]{Lewin2006}. 

\citet{Guver2006} studied the spectral emission of U~Gem in both quiescence and outburst states, using  
the same \textit{XMM-Newton}, \textit{Chandra} HETG/ACIS, and \textit{RXTE} PCA observations 
 present in our sample. They fitted the quiescence spectra with both \textsc{cemekl} (or \textsc{cvmkl}) and \textsc{mkcflow} cooling flow models, finding a maximal temperature $kT_{\rm{max}}$ in the $\sim15-38$~keV range. They report, for the first time, properties of the hard X-ray emission of this source in outburst. In this state, an extra X-ray non-termal component is needed, in addition to a simple mekal-based multi-temperature component,  to obtain an acceptable spectral fit. Though deviations from the continuum are still noticeable, this extra component was better described by a power law rather than other models (such as thermal bremsstrahlung, see Table 4 and Fig.  3 and 4 therein). The authors report $kT_{\rm{max}}=9.49\pm4.1$~keV and $18.44\pm10.5$~keV for the \textsc{cemekl} component in the \textit{Chandra} HETG/ACIS HEG and MEG spectra, respectively; and a power-law component with photon index $\Gamma$ of $0.54\pm0.16$ and $0.64\pm0.06$, respectively. Analyzing the 3-20 keV PCA spectra throughout the 2004 outburst the authors demonstrated  evidence of a transient hard X-ray component -- since in 9 out of 23 spectra the power-law component was not required. They  suggested that this component is driven by a mechanism temporarily present during outburst, which could be either reflection of X-ray coming from either the optically thick plasma  or from an optically thick boundary layer, as well from a  transient magnetosphere present during outburst \citep[see][and reference therein]{Guver2006}. On the other hand, they noticed that spectral features expected by the reflection scenario  such as  the K$_{\alpha}$ fluorescent lines are either very weak or not observed during the outburst state in U~Gem. It is important to stress, however, that reflection scenario is not the only one capable to explain the K$_{\alpha}$ fluorescent line production \citep[see, e.g.,][]{Maiolino2019}. 

\citet{Mukai2003} showed that the \textit{Chandra} spectrum of the DN U~Gem (as well as SS~Cyg and V603~Aql) can be 
described by the multi-temperature bremsstrahlung cooling-flow model \textsc{mkcflow} with $kT_{\rm{max}}=20$ keV. 
\citet{byckling2010} fitted the \textit{XMM-Newton} U~Gem spectrum in quiescence also using the \textsc{mkcflow} 
model and found a maximum temperature $kT_{\rm{max}}=25.82^{+1.98}_{-1.43}$~keV and an equivalent 
width of the Fe 6.4 keV line (EW6.4) of $60^{+33}_{-32}$~eV. Using, on the other hand, a single temperature 
optically thin plasma model (\textsc{mekal}) they found $kT = 0.78^{+0.03}_{-0.01}$~keV and EW6.4 
of $50^{+25}_{-24}$~eV.
\citet{Xu2016} analyzed the spectra of 41 CVs observed with \textit{Suzaku}. They used the single temperature 
model \textsc{apec} to fit the 2--10~keV continuum and 3 Gaussian components (at 6.4~keV, 6.7~keV, and 7.0~keV) 
to fit the Fe emission lines of  U~Gem in quiescence. They found a plasma temperature $kT=16.5^{+4.49}_{-3.31}$~keV; 
and EW 6.4 $=43^{+16}_{-11}$~eV, EW6.7 $=258^{+22}_{-21}$~eV, and EW 7.0 $=178^{+11}_{-11}$~eV.

\subsection{SS Cyg}
\label{sec:SSCyg}

SS~Cyg belongs to the U~Gem type of DN. The system has a WD with mass $M_{WD}$ of $1.19\pm0.02\ M_\odot$, a 
secondary star with mass $M_2$ of $0.704\pm0.002\ M_\odot$ \citep{Friend1990} and K5/5 spectral type \citep{Cat2003}. 
The binary system is located at $166\pm12$~pc \citep{harrison1999,harrison2004},  has an inclination angle $i$ of $37\pm5^{\circ}$ \citep{Cat2003} and a period of 6.603~h \citep{harrison2004}. 
This source is one of the DNe broadly studied in outbursts, as it shows an optical outburst every $\sim$ 50 days, in which 
$m_V$ changes from 12th to 8th magnitude \citep[][see for example, Figure~10.1 therein]{Mauche2001,Lewin2006}. They claim that the  
system shows a reflection component in the quiescence and outburst states, with larger contribution in the outburst 
\citep[][and references therein]{Done1997,Lewin2006}. 
\citet{Mukai2003} found a maximum plasma temperature $kT_{\rm{max}}$ of 80~keV using  fits of  the \textit{Chandra} spectrum of this source in quiescence and the multi-temperature 
\textsc{mkcflow} model.

\citet{Okada2008} analyzed the two \textit{Chandra}/HETG observations present in our sample. They constrained the power law index and  the temperature of the boundary layer in the framework of the \textsc{cemekl} model.

\citet{ishida2009} presented a \textit{Suzaku} XIS and HXD-PIN analysis of this source observed, in November 2005,   
in the quiescence and outburst states. Using the multi-temperature thin plasma model \textsc{cvmekal} (model in XSPEC), they 
found a maximum temperature of the plasma $kT_{\rm{max}}$ of $20.4_{-2.6}^{+4.0}$ (stat) $\pm~3.0$ (sys)~keV in 
quiescence, and of $6.0_{-0.2}^{+1.3}$~keV in outburst. 

\citet{byckling2010} fitted a \textit{SUZAKU} XIS spectrum of SS Cyg in quiescence using the  optically thin plasma 
model \textsc{mekal}  and the cooling model \textsc{mkcflow} (including photoelectric absorption and partial covering 
as well). As a result they obtained a plasma temperature of $10.44_{-0.17}^{+0.16}$~keV and a  maximum temperature of the 
plasma $kT_{\rm{max}}$ of $41.99_{-0.76}^{+1.20}$~keV, respectively. In their fits EW6.4 is equal to 
$75^{+9}_{-4}$~eV and $73^{+6}_{-7}$~eV, respectively.
\citet{Xu2016}, using \textit{Suzaku} data (during an outburst) and the single temperature model \textsc{apec}, fit the 2--10~keV continuum including 3 Gaussian components of the iron emission lines, (at 6.4~keV, 6.7~keV, and 7.0~keV).
As a result they estimated a plasma temperature $kT=8.15_{-0.91}^{+1.23}$~keV; 
and EW6.4 $=67_{-10}^{+11}$~eV, EW6.7 $=415_{-30}^{+41}$~eV, and  EW7.0 $=48^{+15}_{-11}$~eV.

\subsection{VW Hyi}
\label{sec:vwhyi} 
VW~Hyi is a well-studied source. It is located at $54\pm0.1$~pc 
\citep[see][and references therein]{Nakaniwa2019} and  belongs to the SU UMa sub-group of DNe 
\citep{Warner1987}, which undergoes both normal DN outbursts and super-outbursts \citep{Godon2005}. The $M_{WD}$ 
was estimated as $0.63\pm0.15\ M_\odot$ \citep{schoembs1981}\footnote{A gravitational redshift determination done by 
\citet{Sion1997} yielded $M_{WD}= 0.86^{+0.18}_{-0.32}\ M_\odot$}. The secondary star has a mass 
$M_2$ of $0.11\pm0.02\ M_\odot$ \citep{schoembs1981} and it is a M6V star \citep{Godon2005}. The binary system 
has an inclination angle $i$ of $60\pm10^{\circ}$ \citep{schoembs1981}, and an orbital period of 0.074271~days 
\citep[$\sim$1.783~h, which lies below the CV period gap;][]{Cat2003}.  VW~Hyi shows super-humps (which are variations 
in the light curves at a period of a few per cent longer than the orbital period) with a period of 0.07714 days ($\sim$1.85~h). 
VW~Hyi lies along a line of sight with a very low interstellar absorption, $N_H$ $\approx 6 \times 10^{17}$ atoms cm$^{-2}$ 
\citep{Polidan1990}. This low interstellar absorption allows us to observe   this system in almost all wavelength ranges.

The source shows a very low mass accretion rate during quiescence, $\dot{M}\approx10^{-11}\ M_\odot$~yr$^{-1}$, emitting 
significantly in UV \citep[between 35--47\%; see][and references therein]{Godon2005}.
The first X-ray observations of VW~Hyi in quiescence were done in 1984 with \textit{EXOSAT} \citep{Woerd1987}, in which 
persistent hard (1--6~keV) and soft (0.04--1.5~keV) X-ray fluxes were observed. In the spectral analysis, an equally best fit 
was pointed out either using a power-law or a thermal bremsstrahlung model ($kT > 5$~keV) without interstellar absorption 
($N_H<10^{21}$ atoms cm$^{-2}$). 

\citet{Belloni1991} fitted the spectrum of VW Hyi observed in quiescence by \textit{ROSAT} using 
the optically-thin thermal plasma model Raymond-Smith with a single temperature \citep{Raymond1977}. They found a temperature of $2.17\pm0.15$~keV 
for zero interstellar absorption (when absorption was included  and significantly different temperature and flux were not found).
\citet{Hasenkopf2002} analyzed the spectrum of VW Hyi observed with \textit{ASCA}, and using two- and multiple temperature 
thermal plasma models revealed that the resulting temperatures lie mostly between 4 and 10~keV.  

\citet{Pandel2003,pandel2005} analyzed a \textit{XMM-Newton/EPIC} spectrum of VW~Hyi in quiescence (the same observation analyzed 
in this paper). They modeled the spectrum with cooling flow models -- i.e., with multi-temperature thin thermal plasma models 
(\textsc{cemekl, cevmkl} and \textsc{mkcflow}) -- and obtained a maximum temperature $kT_{\rm{max}}$  between 6 and 8~keV. \citet{Pandel2003} attempted to fit the EPIC spectra with 
one or two single-temperature \textsc{mekal} components (as commonly used to fit DNe spectra), but obtained a satisfactory fit only 
when three components of their model were used.

\citet{Nakaniwa2019} have also analysed the same \textit{XMM-Newton/EPIC}  observation present in our sample along with 3 more \textit{SUZAKU/XIS} observations  in quiescence. They report that all 0.2-10 keV spectra are moderately well fitted by the \textsc{cevmkl} and \textsc{vmcflow} cooling flow models, with a maximal temperature of 5-9 keV. Namely, for the \textit{XMM-Newton} observation, they obtained a plasma temperature $kT_{\rm{max}}$ of $5.32_{-0.22}^{+0.23}$~keV and $6.95_{-0.19}^{+0.20}$~keV with the \textsc{cevmkl} and \textsc{vmcflow} models, respectively.

\citet{Xu2016}, using \textit{Suzaku} data and the single temperature model \textsc{apec} to analyze the 2--10~keV continuum, 
and 3 Gaussian components (at 6.4~keV, 6.7~keV, and 7.0~keV)  fit the Fe emission lines, found in VW~Hyi (in the  quiescence state). They found  
a plasma temperature $kT$ of $5.79^{+4.71}_{-2.14}$~keV; and for the emission Fe lines they estimated EW6.4 = 0.01~eV (upper limit), 
EW6.7 = $1392^{+152}_{-172}$~eV, and EW7.0 = $42^{+58}_{-42}$~eV.

\subsection{SS Aur}
\label{sec:SSAur}

SS~Aur belongs to the U~Gem type of DNe with high mass accretion rates $\dot{M}$. The binary system is located at $167^{+10}_{-09}$~pc 
\citep{harrison2004} and has an inclination angle $i$ of $38\pm16^{\circ}$. The system has an orbital period of 0.1828(1)~days 
(which is equivalent to 4.387~h) \citep{shafter1986,byckling2010,harrison2004}. The system hosts a WD with a mass $M_{WD}$ of 
$1.08\pm0.40\ M_\odot$, and a secondary star of $\sim$M1~V type \citep{harrison2000} with a mass $M_2$ of $0.39\pm0.02\ M_\odot$ 
\citep[][and references therein]{Cat2003}.
\citet{byckling2010} fitted a \textit{SUZAKU} XIS spectrum of SS Aur in quiescence using both the optically thin plasma model \textsc{mekal}  
and the cooling model \textsc{mkcflow} (including photoelectric absorption and partial covering as well). They obtained a plasma temperature 
of $6.35\pm0.40$~keV and a  maximum temperature of the plasma $kT_{\rm{max}}$ of $23.47_{-3.02}^{+4.01}$~keV, respectively. In these 
fittings EW6.4 is equal to $73^{+37}_{-36}$~eV and $86^{+52}_{-53}$~eV, respectively.
\citet{Xu2016}, using \textit{Suzaku} data and the single temperature model \textsc{apec} in the 2--10~keV continuum and 3 Gaussian (the Fe emission lines) components (at 6.4~keV, 6.7~keV, and 7.0~keV)  found that  SS~Aur, in the quiescence state,  has the best fit parameters of a plasma temperature of $8.48^{+5.34}_{-2.61}$~keV; and EW6.4 = $47^{+60}_{-29}$~eV, EW6.7 = 325$^{+90}_{-65}$~eV, and EW7.0 = $169^{+57}_{-49}$~eV. 

\section{Data Reduction}
\label{sec:datared}

\subsection*{\textit{XMM-Newton}}
\begin{table*}
\caption{Log of the \textit{XMM-Newton}, \textit{Chandra} and \textit{RXTE} observations}
\label{tab:obs}
\scriptsize
\center
\begin{tabular}{llllllll}
\toprule 
Source    &Observatory &Instrument    &Obs.~ID      & Start Time    &End Time      & Exp. & State\\
    	       &                     &                     &                   & Date (UTC)  & Date (UTC)  & (ks) &\\
\midrule
VW~Hyi  &XMM-Newton &EPIC/PN      &0111970301     &2001-10-19~06:10:05   &2001-10-19~10:38:34  &14.5  &quiescence\\ 
               &Chandra*         &LETG/HRC  &21671               &2018-08-08~00:16:30   & 2018-08-08~03:19:03 &9.67  &outburst\\    
SS~Cyg  &XMM-Newton &EPIC/PN      &0111310201     &2001-06-05~08:14:19   &2001-06-05~11:34:10  &11.8   &quiescence\\
               &Chandra         &LETG/HRC  &1897                 &2001-01-16~21:13:00   &2001-01-17~10:49:56  &47.1  &outburst\\
               &Chandra         &HETG/ACIS &646                   &2000-08-24~10:28:23   &2000-08-25~00:19:30  &47.3  &quiescence\\ 
               &Chandra         &HETG/ACIS &648                   &2000-09-14~21:09:02   &2000-09-15~14:15:05  &59.5  &outburst\\
               &Chandra         &HETG/ACIS &2307                 &2000-09-12~17:00:58   &2000-09-13~03:47:18  &36.6  &outburst\\
               &RXTE             &PCA             &50012-01-01-00 &2000-08-24~13:04:16 &2000-08-24~18:31:44  &13.4  &quiescence\\
               &RXTE             &HEXTE        &50012-01-01-00 &2000-08-24~13:04:16 &2000-08-24~18:31:44  &1.19  &quiescence\\
               &RXTE             &PCA              &10040-01-01-000&1996-10-09~16:47:28 &1996-10-09~23:00:00 &6.10  &outburst\\
               &RXTE             &PCA              &10040-01-01-001&1996-10-09~23:28:32 &1996-10-10~07:25:36 &11.4  &outburst\\
               &RXTE             &PCA              &10040-01-01-00  &1996-10-10~07:25:20 &1996-10-10~11:17:04  &8.67 &outburst\\                                             
U~Gem  &XMM-Newton &EPIC/PN       &0110070401        &2002-04-13~05:35:37 &2002-04-13~11:35:35  &15.1 &quiescence\\
               &Chandra         &LETG/HRC   &3773                   &2002-12-25~19:50:15 &2002-12-26~09:28:39  &47.0 &outburst\\   
               &Chandra         &HETG/ACIS  &647                     &2000-11-29~12:01:20 &2000-11-30~15:13:31  &94.9  &quiescence\\   
               &Chandra         &HETG/ACIS  &3767                   &2002-12-26~09:28:39 &2002-12-27~03:33:37  &61.4 &outburst\\
               &RXTE             &PCA/HEXTE &80011-01-02-00  &2004-03-05~02:14:56  &2004-03-05~09:03:44 &14.5 &outburst\\   
SS~Aur  &XMM-Newton &EPIC/PN       &0502640201       &2008-04-07~08:40:28 &2008-04-07~20:31:58  &31.5  &quiescence\\
               &RXTE             &PCA/HEXTE &30026-03-01-00 &1998-01-24~05:03:12 &1998-01-24~12:12:00  &14.2 &quiescence\\
\bottomrule 
\end{tabular}
\begin{itemize}
\item[*] Observation not considered in our final analysis. 
\end{itemize}
\end{table*}

We analyzed in total four \textit{XMM-Newton} EPIC-pn public observations, 
one observation of each source. See Table \ref{tab:obs} for a log of the observations. 
All \textit{XMM-Newton} observations were taken with the sources in the quiescence state. 

 VW~Hyi, U~Gem and SS~Aur  were observed with the pn camera operating 
in Imaging mode, while SS~Cyg was observed with the pn camera in Timing mode.
All light curves and spectra were extracted through the Science Analysis Software (SAS) 
version 14.0.0. We strictly followed the  recommendations for the pn camera in each mode 
of observation (timing and imaging, respectively).  Following \citet{XMMcal0018}, we 
considered the 0.3$-$15.0 keV spectral energy range for observations taken in Imaging 
mode, and  the 0.7$-$10.0 keV range for the observation taken in Timing mode. 

We checked spectral variability extracting two light curves for each source. 
The two light curve energy ranges were selected accordingly to the observational 
mode: from 0.3 to 4.0~keV and 4.0 to 15~keV for observations taken in Imaging 
mode; and from 0.7 to 4.0~keV and 4.0 to 10.0~keV for observation taken in 
Timing mode. We computed ratios between the light curves in the two energy 
bands (hardness ratio, HR). The HR of SS~Aur showed a time interval of increased 
noise in the end of the exposure, which was discarded in the spectral extraction. 
For all other observations, {a count rate variability was not associated with a significant 
change in the HR or with noise increase. We used therefore the total EPIC-pn exposure time in the spectral extraction of these observations.}

Standard filters were applied during data screening through the \texttt{evselect} task. 
We considered only single and doubles events (using pattern $\leq 4$). \texttt{FLAG==0} 
was used to discard regions of the detector (like border pixels and columns with higher offset) 
for which the pattern type and the total energy is known with significantly lower precision. 

{The observation of SS~Cyg is taken in Timing mode and prior 23rd of May 2012, period in which almost all EPIC-pn exposures are unexpectedly affected by X-ray loading (XRL) – which occurs when source counts contaminates the offset map taken prior the exposure. Because of this we applied the XRL correction to this observation.} 

In all spectral extraction, the background region was selected from a region away from the source. 
The distribution matrix (rmf) and the ancillary (arf) files were created through the \texttt{rmfgen} and 
\texttt{arfgen} tasks. The final spectra were rebinned in order to have at least 25 counts for each 
background-subtracted channel.

\subsection*{\textit{Chandra}}

We used all 8 grating \textit{Chandra} public  observations of our nmCV sample presented to the  preparation time of this paper,
see Table \ref{tab:obs} for a 
log of the observations. VW Hyi was observed only once during an outburst, 
while SS Cyg and U Gem were observed four and three times, respectively, 
in both outburst and quiescence states. SS Aur does not have public observations available in the archive.     
SS Cyg, U Gem and VW Hyi were observed once in LETG/HRC-S mode, while 
SS Cyg and U Gem were observed 3 and 2 times in HETG/ACIS-S mode, respectively. 
Following the standard procedure\footnote{https://cxc.harvard.edu/ciao/threads/gspec.html}, 
we used CIAO v4.11 and the corresponding calibration files to reprocess the data. The 
spectra of each observation were generated with the {\it Chandra\_repro} script. 
We combined the first order spectra together and adopted a minimum signal-to-noise 
value of 10 to group the spectra. We also extracted for each observation two light curves 
in 0.4-4 keV and 4-10 keV bands, and checked for variability. Significant changes in the HR were not 
observed and therefore we used  the total exposure time in the spectral extraction.
The reprocessed spectrum of VW Hyi did not show a spectral component at energies greater 
than 0.5 keV. Therefore,  we did not consider this spectrum in our final spectral analysis.

\subsection*{\textit{RXTE}}
We have analyzed 6 \textit{RXTE} 
observations of three nmCVs:  one of U~Gem, 
one of SS~Aur, and four of SS~Cyg. 
SS~Aur and SS~Cyg Obs. ID 50012-01-01 were 
observed in quiescence. All other observations were taken with the source in outburst state. 
See Table \ref{tab:obs} for a log of the observations.  

All PCA data was analyzed following the standard procedure. Only low
resolution (16{\thinspace}s) light curves were produced together with
spectra in the 2--60{\thinspace}kev band. For each observation,  spectra
and light curves were derived using standard {\it RXTE}/FTOOLS. Briefly, the
data reduction consists of the source{\thinspace}$+${\thinspace}background
production, with subsequent subtraction of the PCA background estimated through 
the  \textsc{pcabackest} tool. The PCA response matrix was built with \textsc{pcarsp}, and a systematic errors (off 0.04{\%}) were added to each individual spectrum.

HEXTE  20--200{\thinspace}keV spectrum extraction \citep{Roths1998} is done using 
comparison, for  each cluster, the spectra from two separate background 
fields that are sampled alternately during observations. This capability
allowed us, for example, in discarding data from cluster B for
SS~Cyg, since one of the background regions for this cluster
(and for that sky position) is contaminated by the nearby source
IGR~J21485$+$4306. HEXTE background lines \citep{Roths1998} were also treated and removed from the final spectrum. Spectra and response matrices were produced following the well-known procedures \citep[see, e.g.,][]{Damico2001}.

\section{Spectral Analysis}
\label{sec:specanalysis}

In order to validate the Comptonization framework in nmCVs and  to obtain  better estimates of 
the physical parameters and spectral (photon) indices in these systems, we analyzed spectra of our sources 
obtained by three different observatories -- \textit{XMM-Newton}, \textit{Chandra} and \textit{RXTE}. 
All analyses were performed using XSPEC astrophysical package 
version 12.8.2. 

The spectral continuum was first modeled by a single thermal Comptonization component, 
\textsc{compTT} model in XSPEC \citep{Titarchuk1994,Hua1995,Titarchuk1995}, considering a plasma cloud of a plane geometry. The physical free parameters of this model are: the temperature  
of the seed photons, $kT_s$,  the electron temperature $kT_e$ and the optical  depth $\tau$ of the Compton cloud. Only when required by the fit, a second Comptonization component 
was added to the total model. Due to the broader spectral energy range of \textit{RXTE}, this second component 
was required only in three out of four \textit{RXTE} spectra of SS Cyg.

In all \textit{XMM-Newton}, \textit{Chandra} and \textit{RXTE} spectra, the residual excess -- i.e., the emission lines  
--  expected and observed in the Fe XXI--XXVI K-shell ($\sim$6.4-7.0 keV) energy range was modeled by up 
to three Gaussian components. Due to the low energy resolution of \textit{RXTE}/PCA, only one broad 
Gaussian component is required to account for the excess caused by these features. 

\subsection*{XMM-Newton Spectral Analysis}
In all spectra, an addition of a photo-electric absorption ($N_H$) and/or a blackbody component (e.g., \textsc{bbody} 
or \textsc{bbodyrad} model) did not improve the fit performed in the $\sim~0.3-15/0.4-10$ keV energy range of 
\textit{XMM-Newton}. 

The spectrum of U~Gem showed three iron emission lines, with centroid energies of: $6.394_{-0.013}^{+0.013}$ keV 
{(at 2.5 sigma of the He-like K$_{\alpha}$ line)}, $6.68_{-0.04}^{+0.04}$ keV (compatible with He-like Fe K$_{\alpha}$ line) 
and $6.972^{+0.023}_{-0.019}$ keV (at 1.2 $\sigma$ from the H-like Fe K$_{\alpha}$ line at 7.0 keV); 
and equivalent widths (EWs) of $53^{+41}_{-25}$ eV, $281^{+100}_{-49}$ eV and $167^{+53}_{-40}$ eV, respectively. 

The spectra of  SS~Cyg and SS~Aur showed two iron emission lines. In SS~Cyg, the lines appear with centroid energies 
of $6.634_{-0.023}^{+0.026}$ keV (compatible with neutral Fe K$_{\alpha}$ line) and $6.966^{+0.030}_{-0.026}$ keV 
(at 1.1 $\sigma$ from the H-like Fe K$_{\alpha}$ line at 7.0 keV); and EWs of {$186_{-27}^{+321}$} eV and $75^{+97}_{-20}$ eV, 
respectively.

In SS~Aur, the lines appear with centroid energies of $6.67_{-0.03}^{+0.04}$ keV (compatible with He-like Fe K$_{\alpha}$ line) 
and $7.00_{-0.05}^{+0.05}$ keV (compatible with H-like Fe K$_{\alpha}$ line); with EWs of $528^{+388}_{-156}$ eV and 
$169^{+99}_{-89}$ eV, respectively. 

The VW~Hyi spectrum is the only one that showed  one emission iron line. This line appears with a Gaussian centroid energy 
of $6.672^{+0.017}_{-0.015}$ keV (at 1.6 $\sigma$ from the He-like Fe K$_{\alpha}$ line); and it is very strong, with EW equal to $995_{-143}^{+152}$ eV, 
which is 2$-$20 times stronger than those observed in other sources. 

\begin{figure}
\centering
\includegraphics[width=8.5cm]{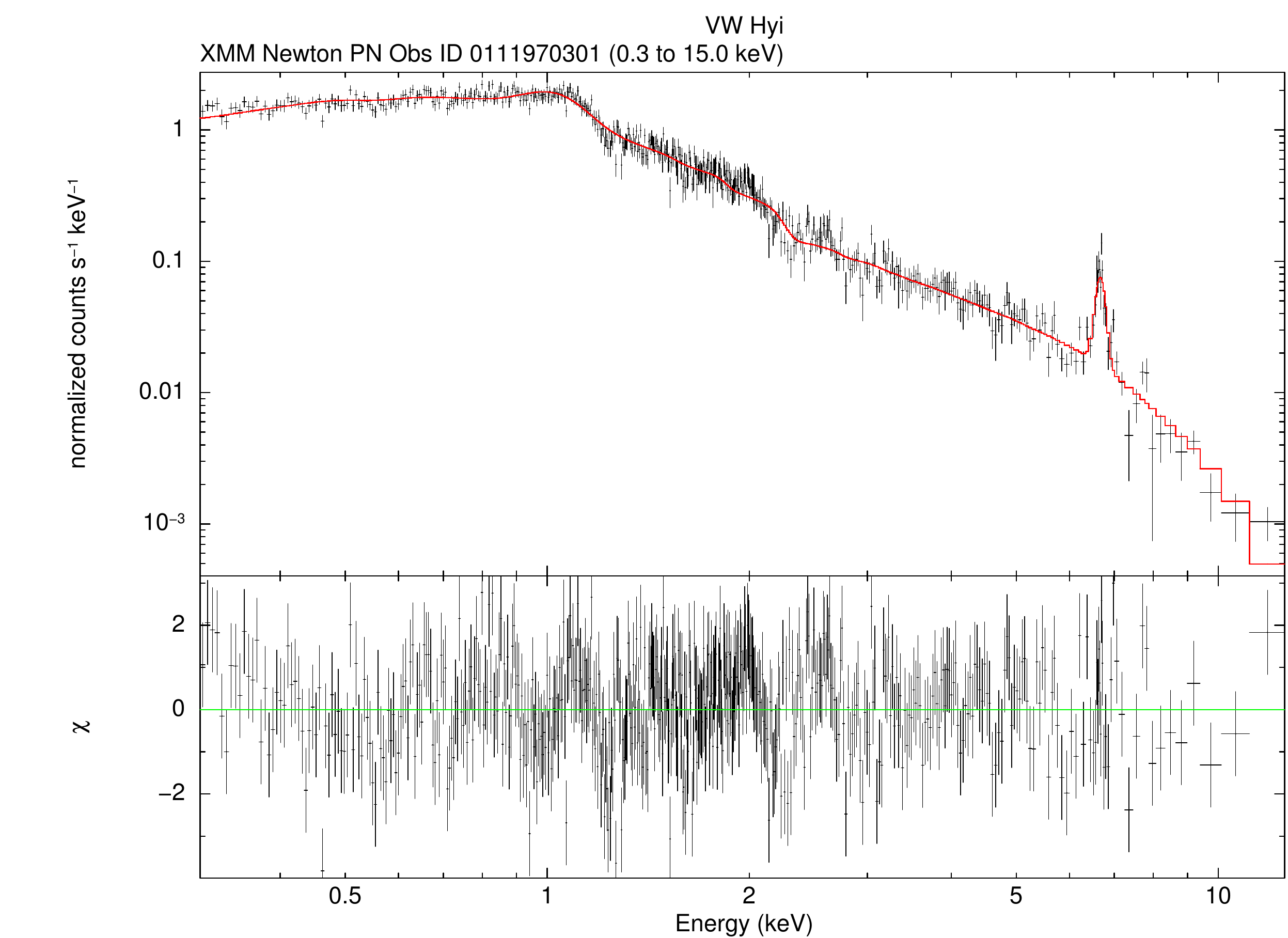}
\includegraphics[width=8.5cm]{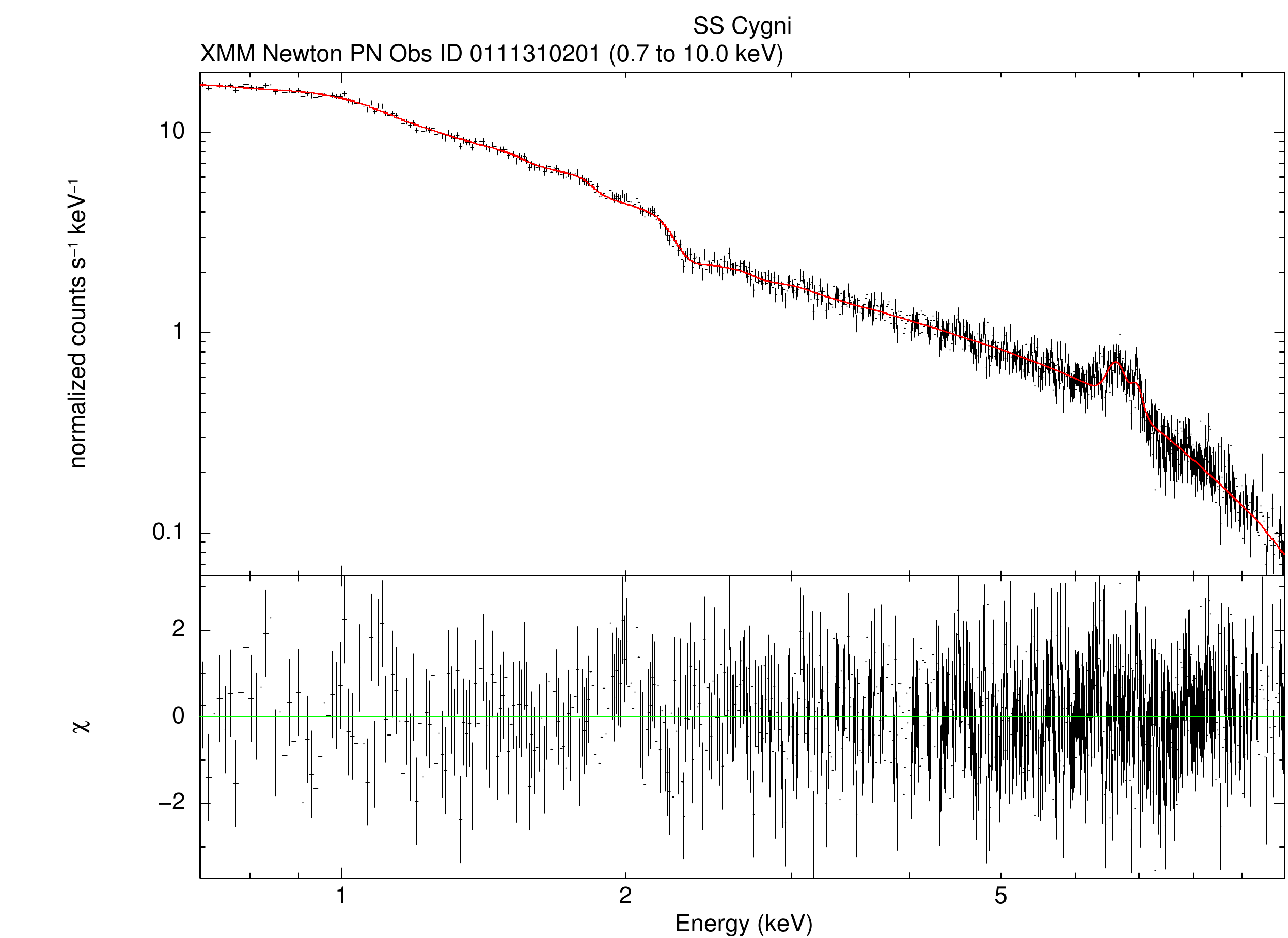}
\includegraphics[width=8.5cm]{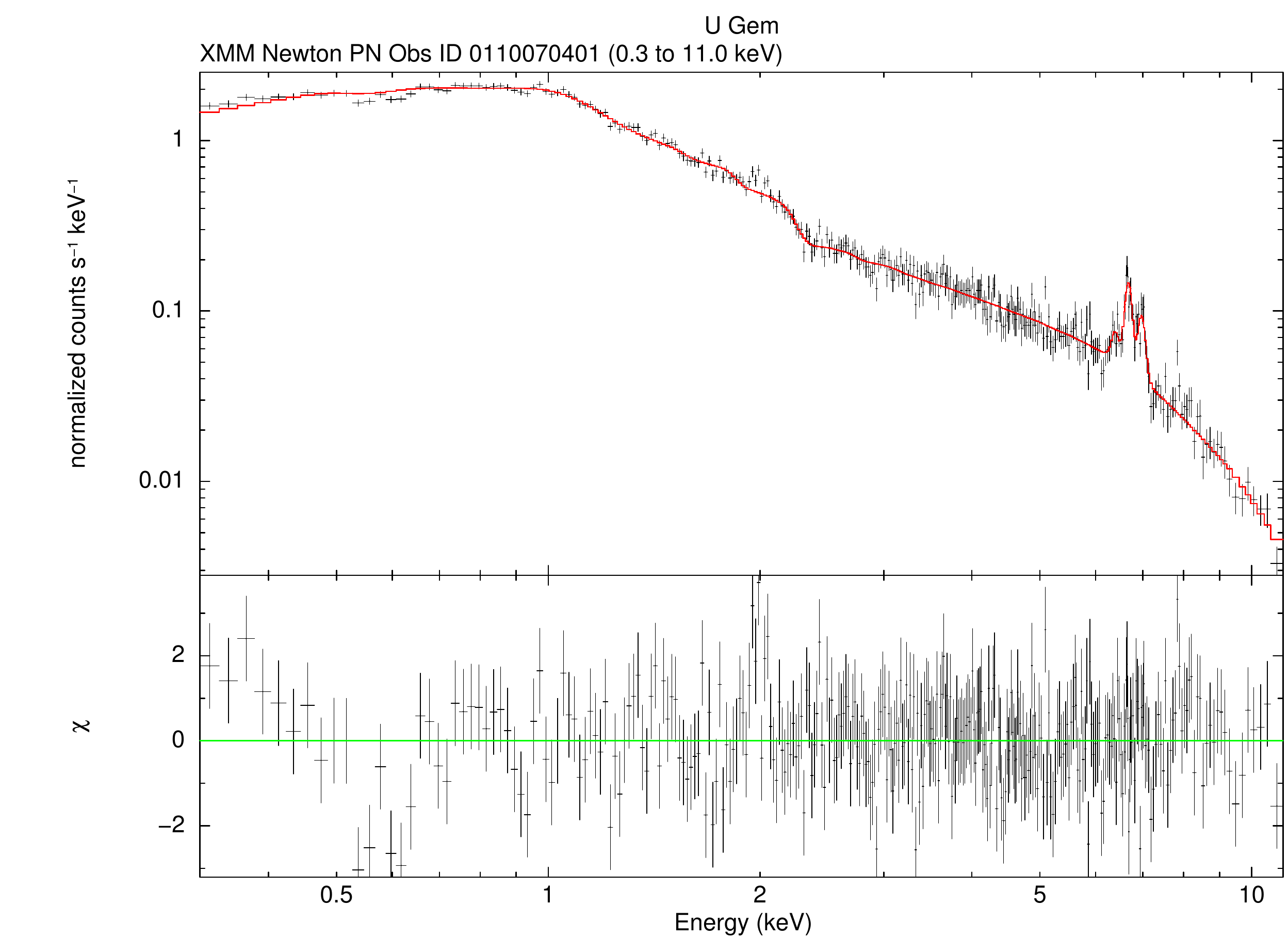}
\includegraphics[width=8.5cm]
{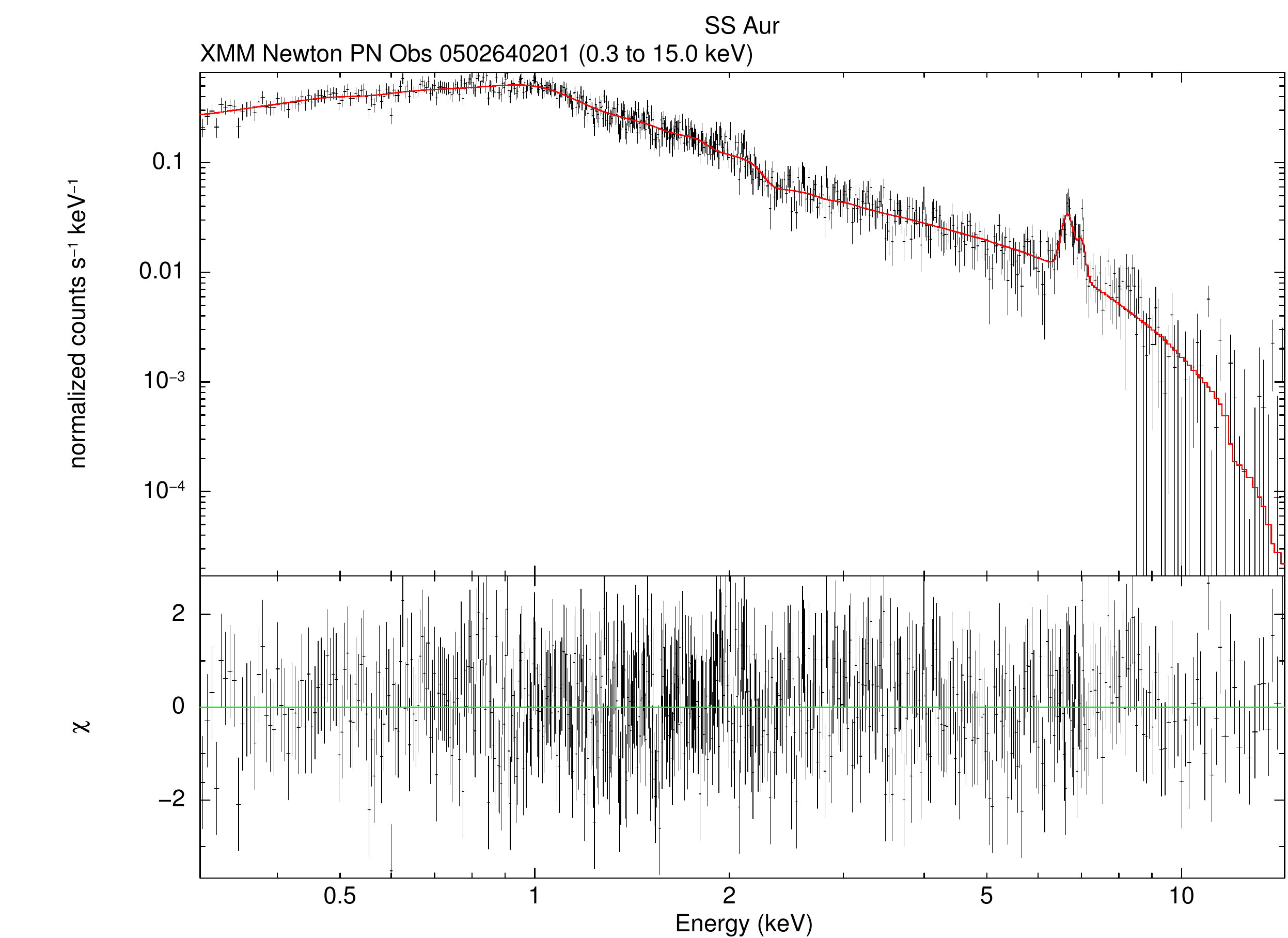}
\caption 
{XMM Newton EPIC pn spectra of the four non-magnetic CVs analyzed. \textit{Left top panel}: spectrum  
   of VW Hyi in the 0.3-15.0 keV energy range, the data (\textit{in black}) and the total model \textsc{compTT+gaussian {+gaussian}} 
  (\textit{solid red line}). \textit{Left lower panel}: spectrum of U Gem in the {0.3-15.0} keV energy range, the data (\textit{in black}) 
   and the best fit model \textsc{compTT+gaussian+gaussian+gaussian {+gaussian}} (\textit{solid red line}). \textit{Right top panel}: spectrum
   of SS Cyg in the 0.7-10.0 keV energy range, the data (\textit{in black}) and the best fit model \textsc{compTT+gaussian+gaussian{+gaussian}} 
    (\textit{solid red line}). \textit{Right lower panel}: spectrum of SS Aur in the 0.3-15 keV energy range, the data (\textit{in black}) 
     and the best fit model \textsc{compTT+gaussian+gaussian {+gaussian}} (\textit{solid red line}). The lower panels show the residuals 
     of the data vs. model
}
\label{fig:specXMM}
\end{figure}

Table \ref{tabT4:GanmCV} shows the best-fit parameters and fit quality found for each \textit{XMM-Newton} observation, 
and Figure \ref{fig:specXMM} shows the best spectral fits. The presence of the Fe Gaussian lines are evident in all fits. Without adding any model to account for the iron line emission, the best-fit 
gives a reduced 
$\chi^{2}_{\rm red}=\chi^2$/degree of freedom (d.o.f)) of 1.54 (801/521) for VW~Hyi, 1.38 (1151/836) 
for SS~Cyg, 1.97 (530/269) for U~Gem and 1.11 (672/604) for SS~Aur. 
When Gaussian components were added to the fit, it leads to a $\chi^{2}_{\rm{red}}$ of 1.30 (672/519) for VW~Hyi, 1.01 
(840/831) for SS~Cyg, 1.15 (366/317) for U~Gem and 0.93 (556/598) for SS~Aur.

In addition to the emission iron lines in the $\sim$ 6.4 - 7.0 keV energy range, we observed in all spectra another strong 
and broad residual excess peaked at $\sim$ 1.01 keV,  with Gaussian centroid in the $\sim$ 0.96 - 1.02 keV energy range 
(see the Gaussian$_0$ component in Table \ref{tabT4:GanmCV}). 
These centroid energies are compatible with resonance lines emitted by Ne X, Fe XVII 
or Fe XXI ions. Table \ref{tab:1lines} summarizes these possible emission lines.

We determined the range of physical parameters given by the \textsc{compTT} model: 

\begin{itemize}
\item The temperature $kT_s$ of the seed photons ranges from 0.056 to 0.174~keV. 

\item The electron temperature, $kT_e$ of the Comptonization cloud ranges from 5.99 to 8.72~keV.

\item The optical depth $\tau$ of the Comptonization cloud ranges from 2.65 to 4.73.
\end{itemize}

\begin{table*}
\caption{Spectral analysis of the \textit{XMM-Newton} observations of our nmCVs sample. Our  model to fit  the spectra: \textsc{comptt+gaussian$_0$+gaussian$_1$+gaussian$_2$+gaussian$_3$}.}
\label{tabT4:GanmCV}
\centering \small
\begin{tabular}{lllllllllllll} 
Source          &                      &               &VW Hyi                                  & SS Cyg                              & U Gem                                      & SS Aur \\ 
\cline{4-7}                                                                                                                            
Component    &Parameter    &Unit& &  & &  \\ 
\toprule
\vspace{1.5mm} 
CompTT & $kT_s$           &keV             &$0.15_{-0.01}^{+0.01}$   &$0.100_{-0.044}^{+0.021}$         &$0.141_{-0.004}^{+0.004}$    &$0.168_{-0.007}^{+0.006}$\\
\vspace{1.5mm} 
             & $kT_e$             &keV             &$8.58_{-0.14}^{+0.14}$         &$7.64_{-0.05}^{+0.05}$         &$6.43_{-0.06}^{+0.06}$          &$6.12_{-0.13}^{+0.13}$\\
\vspace{1.5mm} 
             & $\tau$               &                   &$2.72_{-0.07}^{+0.07}$         &$4.49_{-0.05}^{+0.05}$          &$4.65_{-0.08}^{+0.08}$         &$4.56_{-0.14}^{+0.15}$\\
\vspace{1.5mm} 
Gaussian$_0$ & E$_L$    &keV            &$1.017_{-0.013}^{+0.013}$    &$0.971_{-0.013}^{+0.013}$   &$1.004_{-0.017}^{+0.017}$    &$0.987_{-0.027}^{+0.024}$\\        
\vspace{1.5mm}                                
             & $\sigma$           &keV            &$0.096_{-0.013}^{+0.013}$    &$0.113_{-0.012}^{+0.013}$   &$0.136_{-0.014}^{+0.016}$    &$0.118_{-0.027}^{+0.029}$\\
\vspace{1.5mm} 
             & EW                    &eV              &$150^{+20}_{-19}$                 &$49^{+91}_{-15}$                  &$103^{+20}_{-20}$                &$105^{+25}_{-24}$\\
\vspace{1.5mm} 
Gaussian$_1$ & E$_L$    &keV            &                                               &   &$6.394_{-0.013}^{+0.013}$    &\\        
\vspace{1.5mm}                                
             & $\sigma$           &keV            &                                               &         & $1.3\times 10^{-2}$$^{a}$    &\\
\vspace{1.5mm}
             & EW                    &eV             &                                               &              &$53^{+41}_{-25}$                  &\\
\vspace{1.5mm}
Gaussian$_2$ &E$_L$     &keV           &$6.672^{+0.017}_{-0.015}$     &  $6.634_{-0.023}^{+0.026}$          &$6.68_{-0.04}^{+0.04}$         &$6.67_{-0.03}^{+0.04}$\\        
\vspace{1.5mm}
             & $\sigma$           &keV           &$1.27^{+0.02}_{-0.02}\times10^{-2}$ & $0.12_{-0.05}^{+0.05}$                                   &$1.3\times10^{-2}$$^{a}$      &$8_{-6}^{+6}\times10^{-2}$\\
\vspace{1.5mm}
             & EW                    &eV             &$995_{-143}^{+152}$            & $186_{-27}^{+322}$           &$281^{+100}_{-49}$               &$528^{+388}_{-156}$ \\
\vspace{1.5mm} 
Gaussian$_3$ & E$_L$    &keV           &                                              &$6.966^{+0.030}_{-0.026}$    &$6.972^{+0.023}_{-0.019}$    &$7.00_{-0.05}^{+0.05}$\\        
\vspace{1.5mm}
             & $\sigma$           &keV           &                                              &$5\times10^{-2}$$^{b}$          &$2.3\times10^{-2}$$^{a}$      &$1.5\times10^{-3}$$^{a}$ \\
\vspace{1.5mm}
             & EW                    &eV             &                                              &$75^{+97}_{-20}$                  &$167^{+53}_{-40}$                 &$169^{+99}_{-89}$\\
\vspace{1.5mm} 
Fit quality  &$\chi^2$/d.o.f &                 & 672/519                                & 840/831                                & 366/317                                 & 556/598 \\
             &$\chi^2_{\rm{red}}$&              & 1.30                                     & 1.01                                      & 1.15                                       & 0.93 \\
Energy range &                & keV           &$0.3-15.0$                            &$0.7-10.0$                             & $0.3-15.0$                             &$0.3-15$ \\ 
\bottomrule 
\end{tabular}
\begin{itemize}
\item[ ] Uncertainties at 90\% confidence level.
\item[a] Parameter pegged at hard limit.
\item[b] Parameter frozen.
\end{itemize}
\end{table*}


\begin{table*}
\centering\small

\caption{Emission lines compatible with the line centroid energy of the Gaussian component fit to the broad 
residual excess (in the 0.8 to 1.2 keV energy range) present in the soft X-ray spectra of all nmCVs analyzed. }
\label{tab:1lines}
\small
\begin{tabular}{llllll} \hline
\toprule 
Ion & Transition & Energy (keV) & $\lambda({A^\circ})$   & DNe \\
\midrule
Ne X &1s - 2p &1.022 &12.131 &VW Hyi, U Gem\\
Fe XVII &$2p^{6}~^{1}S - 2p^{5}4s^{1}P$ &0.976 &12.703 & SS Cyg, SS Aur\\
	     &$2p^{6}~^{1}S - 2p^{5}4d^{1}P$ &1.023 &12.119 & VW Hyi \\
Fe XXI  &$2s^{2}2p^{2}~^{3}P - 2s2p^{2}3p^{3}P$ &0.992 &12.498 & SS Aur, U Gem\\
             &$2p^{2}~^{3}P - 2p3d^{3}D$ &1.008 &12.300 & VW Hyi, U Gem, SS Aur\\
\bottomrule 
\hline
\end{tabular}
\end{table*}

\subsection*{Chandra Spectral Analysis}

The residual broad excess around 1 keV observed in all \textit{XMM-Newton} spectra was also observed in 
three out of five HETG/ACIS  \textit{Chandra} spectra analysed in this paper.  This feature and the narrow 
emission lines (narrow excesses $\gtrsim 2 \sigma$) present in the Chandra spectra were modeled by 
simple Gaussian components.
U Gem has one HETG/ACIS observation in quiescence and one in outburst state, whereas SS Cyg was observed once in 
quiescence and twice in outburst state. 

Fitting the continuum with only one \textsc{compTT} component without adding any component to account 
for the emission lines, the best-fit gives a $\chi^{2}_{\rm red}$ of 1.85 (131/71) and 2.57 (321/125) for U~Gem 
Obs. ID 647 and 3767, respectively; and 1.24 (589/478), 2.5 (421/168), and 1.57 (143/91) for SS~Cyg Obs. 
ID 646, 648, and 2307, respectively.  When Gaussian components were added to the fit, it leads to a 
$\chi^{2}_{red}$ of {1.03 (67/65) and 1.01 (114/113)} for U~Gem Obs. ID 647 and 3767, respectively; 
and {0.97 (453/467), 0.80 (121/151), and 0.64 (52/81)} for SS~Cyg Obs. ID 646, 648, and 2307, respectively. 

\begin{figure*}
\center
  \includegraphics[width=13cm]{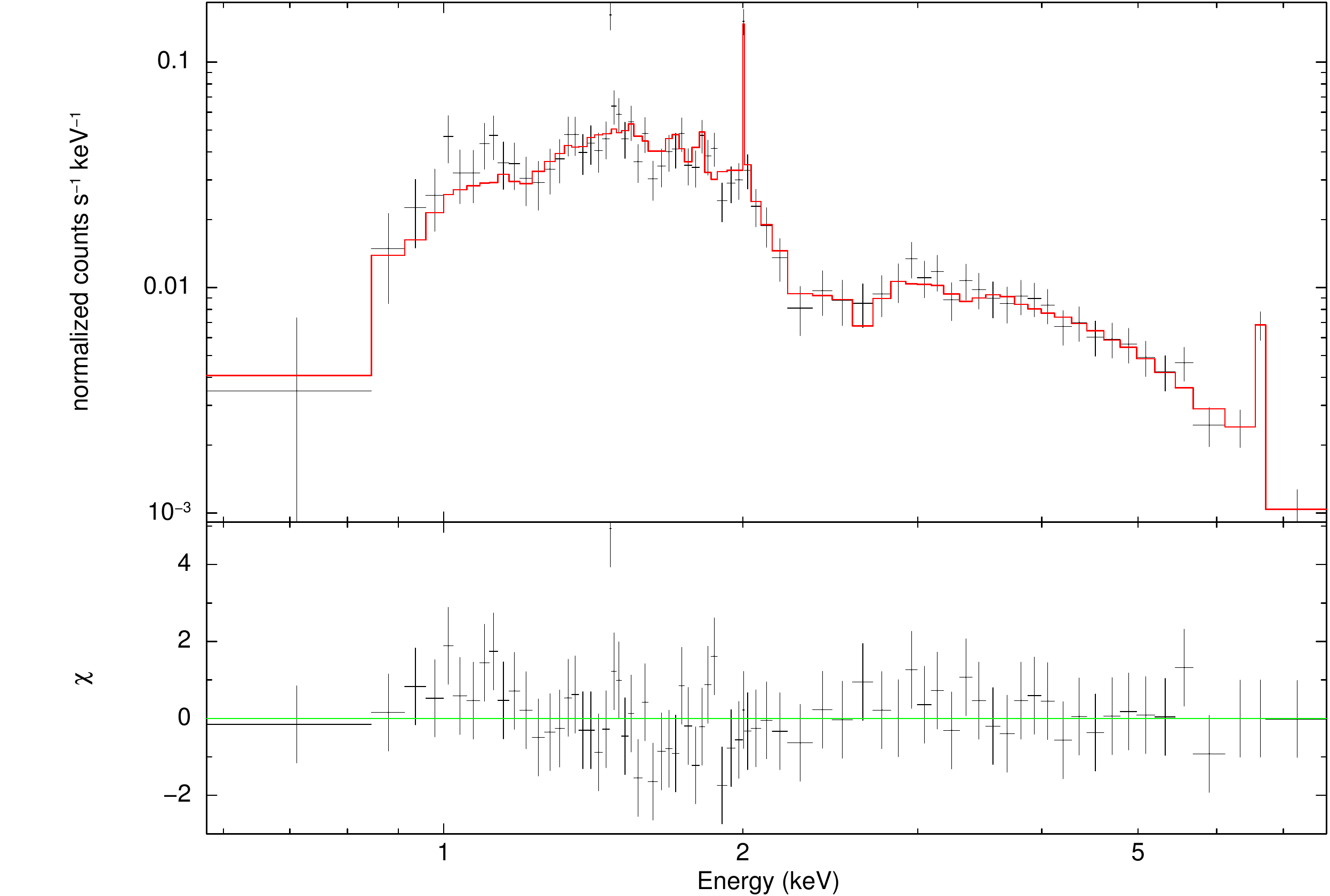}
  \includegraphics[width=13cm]{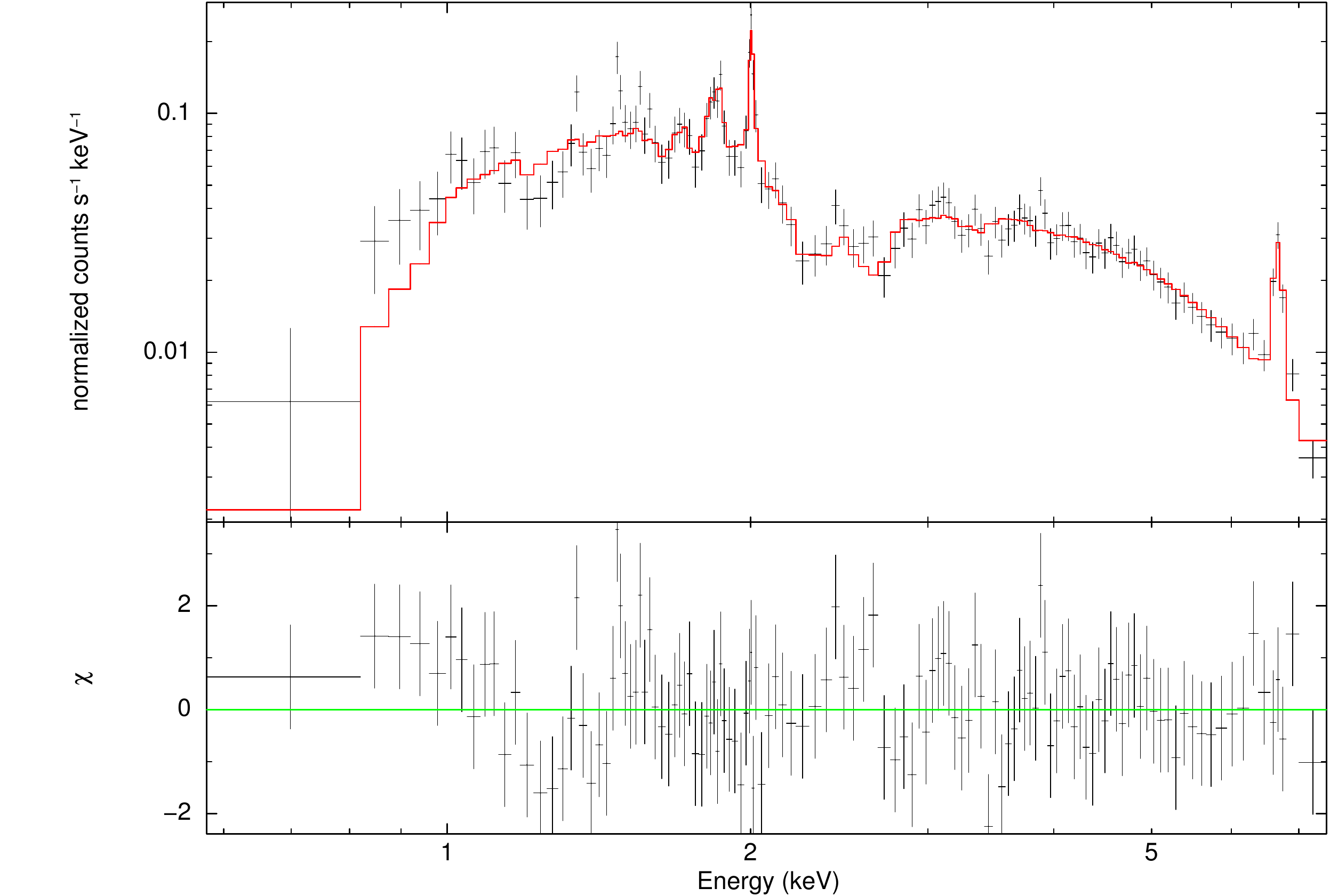}
     \caption{Chandra  HETG/ACIS spectra of U~Gem in the 0.4-10 keV energy range. \textit{Upper  panel}: spectral fit to 
     Obs. ID 647 (see Table \ref{tab:UGemChandraCOMPTT}, column 4), the data (\textit{in black}) and the total model 
     \textsc{compTT+gaussian+gaussian} (\textit{solid red line}). 
     \textit{Bottom panel}: spectral fit to Obs. ID 3767 {(see Table \ref{tab:UGemChandraCOMPTT}, column 5)}, the data (\textit{in black}) and the best fit model 
     \textsc{compTT+gaussian+gaussian+gaussian
     {+gaussian}} components (\textit{solid red line}). The lower panels show the 
     residuals of the data vs. model.}
     \label{fig:UGemspec}  
\end{figure*}

\begin{figure*}
\center
   \includegraphics[width=8.8cm]{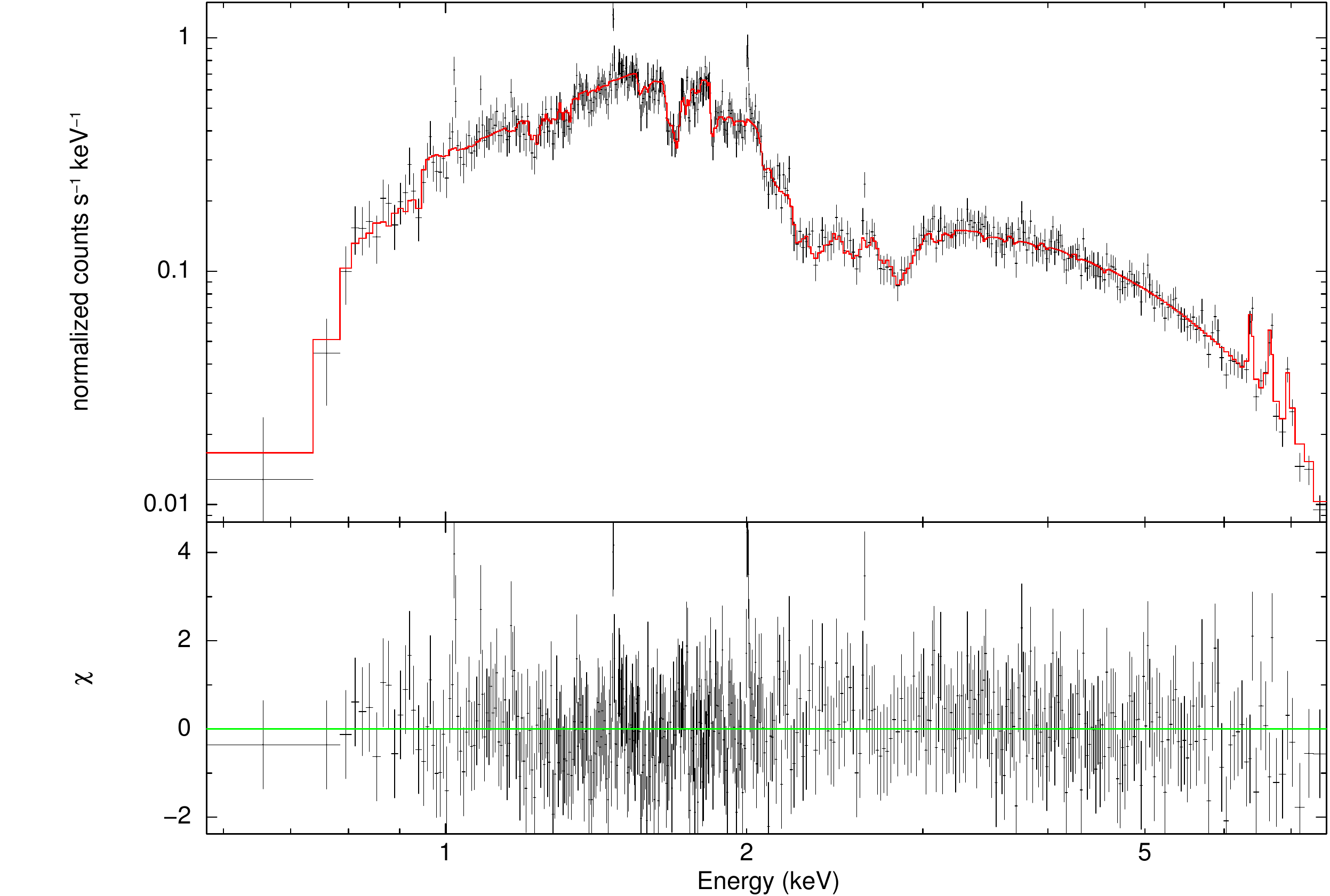}
   \includegraphics[width=8.8cm]{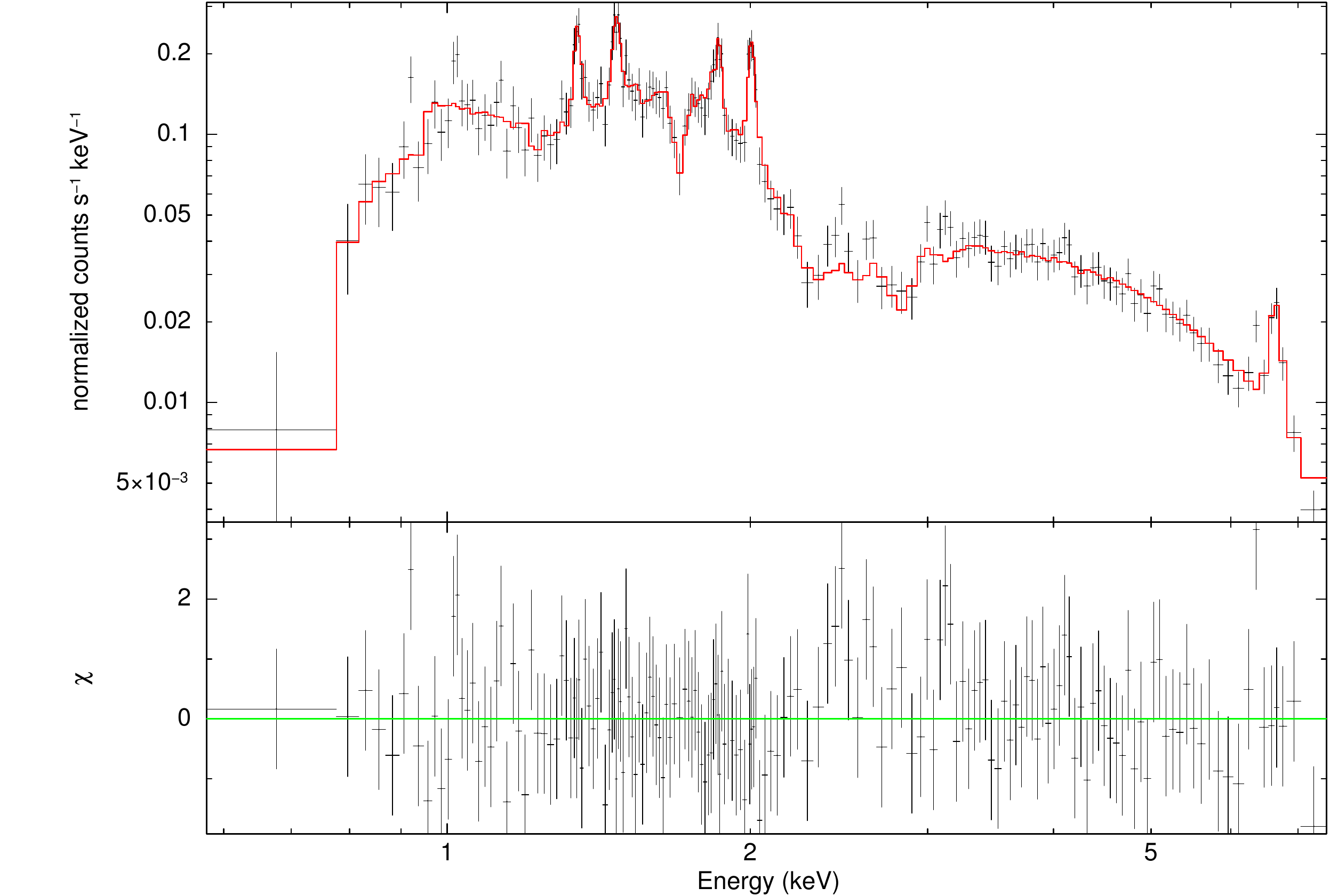}
  \includegraphics[width=8.8cm]{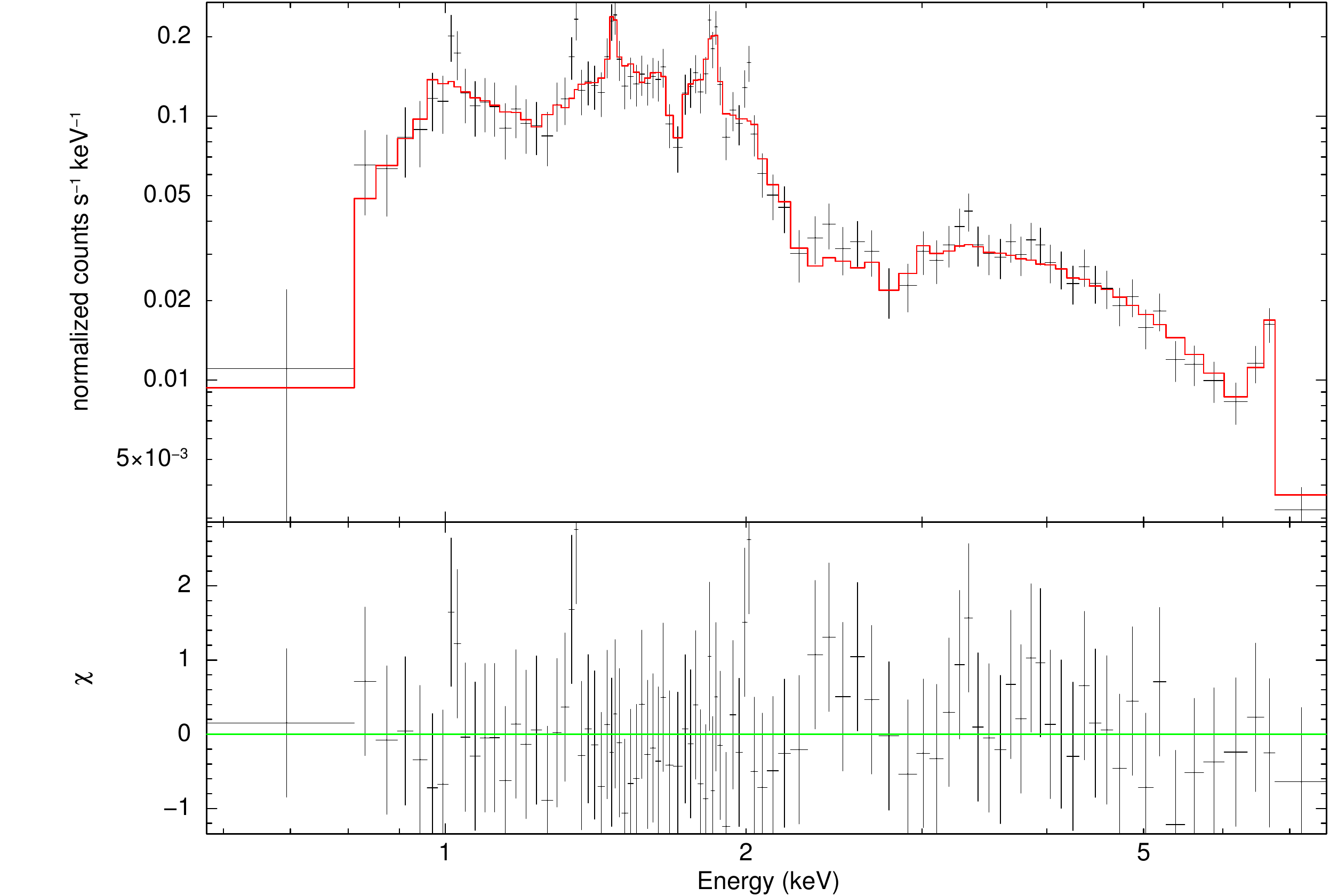}
     \caption{Chandra  HETG/ACIS spectra of SS~Cyg in the 0.4-10 keV energy range. \textit{Left top panel}: 
    spectral fit to Obs. ID 646 {(see Table \ref{tabT4:SSCOMPTT}, column 4)}, the data (\textit{in black}) and the total model 
    \textsc{compTT+gaussian+gaussian+gaussian} 
    (\textit{solid red line}). \textit{Right top panel}: spectral fit to 
     Obs. ID 648 {(see Table \ref{tabT4:SSCOMPTT}, column 5)}, the data (\textit{in black}) and the best fit model 
     \textsc{compTT+gaussian+gaussian+gaussian+gaussian {+gaussian+gaussian}}
     (\textit{solid red line}). 
     \textit{Lower panel}: spectral fit to Obs. ID 2307 {(see Table \ref{tabT4:SSCOMPTT}, column 6)}, the data (\textit{in black})  and the total model \textsc{compTT+gaussian+gaussian+gaussian+gaussian} (\textit{solid red line}). 
     The lower panels in these plots show the residuals of the data vs. model.}
 \label{fig:SSspec}  
\end{figure*}

Table \ref{tab:UGemChandraCOMPTT}  and \ref{tabT4:SSCOMPTT} show the best-fit parameters and fit 
quality for each spectrum of U~Gem and SS~Cyg, respectively. Figures \ref{fig:UGemspec}$-$\ref{fig:SSspec} 
show spectral fits. 

 In U~Gem, the values of the temperature kT$_e$ and $\tau$ parameters, agree at 90$\%$ confidence level, 
 in the two observations:  $\tau$ is $\sim$ 5, 
 and kT$_e$ is equal to $5.0_{-0.5}^{+0.5}$ keV and $6_{-2}^{+5}$ keV in Obs. 647 (in quiescence) and 3767 (in outburst), respectively. 
 On the other hand, a value of the temperature kT$_s$ 
 of the seed photons in Obs. 3767 ($0.66_{-0.10}^{+0.12}$ keV) appears $\sim$ 
 {8} times greater than the 
 temperature in Obs. 647 ($0.66_{-0.10}^{+0.12}$ keV). For the two observations, the total model led to 
 $\chi^2_{red}$ of 1.0  (see Table \ref{tab:UGemChandraCOMPTT}).
\begin{table*}
\centering\small
\caption{Spectral analysis of the \textit{Chandra} HETG/ACIS observations of U~Gem, using pure thermal \textsc{compTT} and Gaussian components.}
\label{tab:UGemChandraCOMPTT}
\small
\begin{tabular}{llllll}                                                                                                         
Component    &Parameter    &Unit       &Obs. ID 647                              & Obs. ID 3767 \\ 
\toprule
\vspace{1.5mm} 
CompTT & $kT_s$              &keV          &$0.087_{-0.010}^{+0.004}$      & $0.66_{-0.10}^{+0.12}$\\
\vspace{1.5mm} 
             & $kT_e$               &keV           &$5.0_{-0.5}^{+0.5}$                  & $6_{-2}^{+5}$\\
\vspace{1.5mm} 
             & $\tau$                 &                 &$5.1_{-0.3}^{+0.3}$                  &$5_{-1}^{+1}$\\
\vspace{1.5mm}               
Gaussian$_0$ & E$_L$    &keV          &$6.63_{-0.07}^{+0.08}$               & $6.68_{-0.02}^{+0.02}$\\        
\vspace{1.5mm}                                
             & $\sigma$           &keV          &$0.06_{{-a}}^{+0.04}$                  & $0.06_{-0.02}^{+0.02}$\\
\vspace{1.5mm} 
Gaussian$_1$ & E$_L$    &keV          & $2.006^{+0.004}_{-0.001}$        & $2.004_{-0.005}^{+0.001}$\\        
\vspace{1.5mm}                                
             & $\sigma$           &keV          &$2^{+1}_{{-a}}\times10^{-3}$       & $9_{-3}^{+3}\times10^{-3}$\\
\vspace{1.5mm}
Gaussian$_2$ & E$_L$    &keV          &                                                    & $1.86_{-0.01}^{+0.01}$\\        
\vspace{1.5mm}                                
             & $\sigma$           &keV          &                                                    & $1.5_{-0.4}^{+0.5}\times10^{-2}$\\
\vspace{1.5mm} 
Gaussian$_3$ & E$_L$    &keV          &                                                     & $1.00^{b}$\\        
\vspace{1.5mm}                                
             & $\sigma$           &keV          &                                                     & $0.24_{-0.06}^{+0.07}$\\
\vspace{1.5mm} 
Fit quality  &$\chi^2$/d.o.f &                 & 67/65                                          & 114/113\\
             &$\chi^2_{red}$   &                  & 1.03                                            & 1.01\\
Energy range &                & keV           &$0.4-10$                                      &$0.4-10$\\   
\bottomrule 
\end{tabular}
\begin{itemize}
\item[ ] Uncertainties at 90\% confidence level.
\item[a] Parameter pegged at hard limit.
\item[b] Parameter frozen.
\end{itemize}
\end{table*}

In SS~Cyg, independently of the source state, the best-fits were found in general by a \textsc{compTT} 
component with kT$_s$ of  0.10 keV, kT$_e$ of $\sim$ 5 keV, and $\tau$ of $\sim$ 6.  It is important 
to stress that initially all parameters were considered as a free one, however, the electron temperature  kT$_e$ 
appears not well constrained by the fits --  it  either assumed values greater 
than the upper limit given by the \textit{Chandra} effective energy band (8 keV), or lower than the minimum 
value acceptable by the Comptonization model (5 keV, see \citet{Hua1995}). Therefore, this parameter 
was kept fixed in the three fits. The spectrum of Obs. 648 is the only one which shows a very low seed 
photon temperature kT$_s$ of $0.020_{-0.003}^{+0.004}$ keV -- the presence of many lines in the soft 
energy band of this spectrum may affect the value of this parameter. We obtained a $\chi^2_{red}$ 
of $\sim$ 1.0 for the fit to Obs. 646, and lower than 1.0 for Obs. 648 and 2307 (see Table \ref{tabT4:SSCOMPTT}).

The LETG/HRC spectra of SS Cyg and U~Gem were not successfully described by only one thermal Comptonization component. 
Moreover, a presence of many lines in the 0.07-2 keV energy band  makes a problem to  satisfactory  fit the data. For example, a good fit in terms of $\chi^2$-statistic ($\chi^2_{red} < 2$) is obtained when the soft energy band 
(E $<$  2.5 keV in SS~Cyg and E $<$ 1.5 keV in U~Gem, respectively) is not taken into account. In this case, fitting SS~Cyg and U~Gem 
spectra with a thermal \textsc{compTB} component  \citep{Farinelli2008} and two Gaussian components --  
with centroid energies at 1.86 and 2.005 keV -- led to $\chi^2_{red}$ of 1.34 (16/12) and 1.33 (47.9/36), correspondingly.

\subsection*{Emission lines in the HETG/ACIS spectra}

The broad residual excess peaked around $\sim$ 1 keV, observed in all \textit{XMM-Newton} Epic-pn spectra, was also 
observed in the \textit{Chandra} spectra of U Gem Obs. ID 3767 in outburst, and SS~Cyg Obs. ID 648 and 2307 (in outburst). 
This broad excess could be an emission from Fe XVII, Fe XXI, or Ne X ions (see Table \ref{tab:1lines}). However, the fit of this 
feature with a Gaussian component led to lower centroid energies. In U~Gem Obs. ID 3767, the Gaussian line energy is frozen at 
1.0 keV, with $\sigma$ equal 
to 
{$0.24 _{-0.06}^{+0.07}$} keV (see Table \ref{tab:UGemChandraCOMPTT}). In SS~Cyg 
Obs. ID 648 and 2307, it is equal to $0.87_{-0.04}^{+0.03}$ keV and $0.93_{-0.05}^{+0.04}$ keV, respectively; 
with $\sigma$ equal to $0.15 _{-0.02}^{+0.03}$ keV and {$0.10_{-0.03}^{+0.04}$} keV, respectively (see Table \ref{tabT4:SSCOMPTT}).
Table \ref{tab:ChandraLines} summarizes the possible resonance narrow emission lines observed in the \textit{Chandra} 
HETG/ACIS spectra. In U~Gem (see Table \ref{tab:UGemChandraCOMPTT}), the spectrum of Obs. ID 647 shows two 
emission lines, with centroid energies at $6.63_{-0.07}^{+0.08}$ keV (compatible with He-like K$_{\alpha}$ Fe line at 6.7 keV) 
and $2.006^{+0.004}_{-0.001}$ keV (compatible with Si XIV, at 2.007 keV). The spectrum Obs. ID 3767 shows three emission 
lines, with centroid energies at $6.68_{-0.02}^{+0.02}$ keV (compatible with He-like K$_{\alpha}$ Fe line), 
$2.004_{-0.005}^{+0.001}$ keV (at 3$\sigma$ of Si XIV line), and $1.86_{-0.01}^{+0.01}$ keV (compatible with Si XIII, at 1.865 keV).

In SS~Cyg (see Table \ref{tabT4:SSCOMPTT}), the spectrum of Obs. 646 shows only the three iron lines in the 6.4-7.0 keV 
energy range: at $6.39_{-0.01}^{+0.01}$ keV (compatible with neutral K$_{\alpha}$ Fe line), $6.67^{+0.01}_{-0.02}$ keV 
(at 3$\sigma$ of the He-like K$_{\alpha}$ Fe line),  and $6.97_{-0.03}^{+0.01}$ (at 3$\sigma$ of the H-like K$_{\alpha}$ Fe line). 
The spectrum of Obs. 648 shows in total six emission lines, with centroid energies at: {$0.87_{-0.04}^{+0.03}$} keV (compatible 
with emission from either Ca XVIII, Fe XVII, Ni XIX, or Ni XVIII; see Table \ref{tab:ChandraLines}), $1.346_{-0.004}^{+0.004}$ keV 
(compatible with emission from Mg XI at 1.343, {and 1.5$ \sigma$ from the 1.352 keV line}), $1.473_{-0.005}^{+0.004}$ (compatible with emission from Mg XII, at 1.473),  
$1.859_{-0.004}^{+0.003}$ keV ({at 2 sigma of} Si XIII emission), $2.005_{-0.004}^{+0.003}$ keV (compatible with Si XIV emission), 
and $6.65_{-0.04}^{+0.03}$ keV ({at 1.7 $\sigma$ from the He-like K$_{\alpha}$ Fe line}). The spectrum of Obs. 2307 shows in total four emission 
lines, with centroid energies at: $0.93_{-0.05}^{+0.04}$ keV ({compatible with Ca XVIII, Ni XIX, and Ne IX}), $1.472_{-0.007}^{+0.008}$ keV(compatible with Mg XII), 
$1.857_{-0.005}^{+0.005}$ keV ({at 1.6 $\sigma$ from the Si XIII line}), and $6.61_{-0.01}^{+0.01}$ keV ({ at 1.7 $\sigma$ from the} Fe XXII K$_{\alpha}$ 
line at 6.627 keV), see Table \ref{tab:ChandraLines}. 

\begin{table*}
\centering\small
\caption{Spectral analysis of the \textit{Chandra} HETG/ACIS observation of SS~Cyg, using thermal \textsc{compTT} and Gaussian components.}
\label{tabT4:SSCOMPTT}
\small
\begin{tabular}{llllll}                                                                                                                       
Component    &Parameter    &Unit       &Obs. ID 646  & Obs. ID 648  & Obs. ID 2307\\ 
\toprule
\vspace{1.5mm} 
CompTT & $kT_s$         &keV           &$0.10_{{-a}}^{+0.05}$           &$0.020_{-0.003}^{+0.004}$               &$0.11_{-0.01}^{+0.01}$ \\
\vspace{1.5mm} 
             & $kT_e$          &keV           &$5.0^{b}$                               &$5.0^{b}$                                           &$5.0^{b}$ \\
\vspace{1.5mm} 
             & $\tau$            &                 &$5.9_{-0.1}^{+0.1}$                &$7.2_{-0.3}^{+0.3}$                           &$5.7_{-0.3}^{+0.3}$\\
\vspace{1.5mm} 
Gaussian$_0$ & E$_L$   &keV          &                                               &$0.87_{-0.04}^{+0.03}$                     &$0.93_{-0.05}^{+0.04}$\\        
\vspace{1.5mm}                                
             & $\sigma$          &keV          &                                               &$0.15 _{-0.02}^{+0.03}$                    &$0.10 _{-0.03}^{+0.04}$\\
\vspace{1.5mm}
Gaussian$_1$ & E$_L$   &keV          &                                               &$1.346_{-0.004}^{+0.004}$                       &\\        
\vspace{1.5mm}                                
             & $\sigma$          &keV          &                                               &$1.09_{-0.3}^{+0.6}\times10^{-2}$$^{a}$   &\\
\vspace{1.5mm}
Gaussian$_2$ & E$_L$   &keV          &                                                &$1.473_{-0.005}^{+0.004}$                      &$1.472_{-0.007}^{+0.008}$\\        
\vspace{1.5mm}                                
             & $\sigma$          &keV          &                                                &$1.2_{-0.3}^{+0.4}\times10^{-2}$$^{a}$   &$1.0_{-0.6}^{+0.8}\times10^{-2}$\\
\vspace{1.5mm}
Gaussian$_3$ & E$_L$   &keV          &                                                &$1.859_{-0.004}^{+0.003}$                     &$1.857_{-0.005}^{+0.005}$\\        
\vspace{1.5mm}                                
             & $\sigma$          &keV          &                                                &$1.3_{-0.2}^{+0.3}\times10^{-2}$            &$1.4_{-0.3}^{+0.4}\times10^{-2}$\\
\vspace{1.5mm}
Gaussian$_4$ & E$_L$   &keV          &                                                &$2.005_{-0.004}^{+0.003}$                     &\\        
\vspace{1.5mm}                                
             & $\sigma$          &keV          &                                                &$1.3_{-0.3}^{+0.3}\times10^{-2}$$^{a}$  &\\
\vspace{1.5mm} 
Gaussian$_5$ & E$_L$    &keV         &$6.39_{-0.01}^{+0.01}$           & $6.65_{-0.04}^{+0.03}$                          &$6.61_{-0.01}^{+0.01}$ \\        
\vspace{1.5mm}                                
             & $\sigma$           &keV         &$0.02^{b}$                              &$0.08_{-0.09}^{+0.08}$                            &$0.02^{b}$\\
\vspace{1.5mm} 
Gaussian$_6$ & E$_L$   &keV          & $6.67^{+0.01}_{-0.02}$         &                                                                 & \\        
\vspace{1.5mm}                                  
             & $\sigma$          &keV          &$0.03^{b}$                             &                                                                  &\\
\vspace{1.5mm}
Gaussian$_7$ & E$_L$   &keV          & $6.97_{-0.03}^{+0.01}$         &                                                                 &\\        
\vspace{1.5mm}                                
             & $\sigma$          &keV          & $0.02^{b}$                             &                                                                &\\
\vspace{1.5mm}
Fit quality  &$\chi^2$/d.o.f &               & 453/467                                 & 121/151                                                  &52/81\\
             &$\chi^2_{red}$    &                & 0.97                                      &0.80                                                        &0.64\\
Energy range &                 & keV         &$0.4-8.0$                               &$0.4-10$                                                &$0.4-10$\\  
\bottomrule 
\end{tabular}
\begin{itemize}
\item[ ] Uncertainties at 90\% confidence level.
\item[a] Parameter pegged at hard limit.
\item[b] Parameter frozen.
\end{itemize}
\end{table*}

\begin{table*}
\centering\small
\caption{X-ray lines, other than the iron lines expected in the 6.4-7.0 keV energy range, compatible with the centroid energy of a Gaussian component present in the \textit{Chandra} HETG/ACIS spectra.}
\label{tab:ChandraLines}
\small
\begin{tabular}{llllll}
\toprule 
Ion & Transition & Energy (keV) & $\lambda({A^\circ})$ & DNe & Obs.ID\\ 
\midrule
Ca XVIII &$2s-4p$                                     &0.886 &13.993 &SS~Cyg &648, 
{2307}\\
Fe XVII  &$2s^{6}-2p^{5}3d^{1}P$                      &0.826 &15.010 &SS~Cyg &648\\    
Ni XIX    &$2p^{6}-2p^{5}3s^{3}P$                     &0.884 &14.025 &SS~Cyg &648, 
{2307}\\
Ni XVIII  &$2s^{6}-3s^{2}S-2p^{5}3s^{2}~^{2}P$ &0.879 &14.105 &SS~Cyg &648\\
Ne IX     &$1s^{2}~^{1}S-1s2p^{1}P$                   &0.921 &13.461 &SS~Cyg &2307\\
Mg XI     &$1s^{2}~^{1}S-1s2p^{3}P$                   &1.343 &9.232 &SS~Cyg &648\\
Mg XI     &$1s^{2}~^{1}S-1s2p^{1}P$                   &1.352 &9.170 &SS~Cyg &648\\
Mg XII    &$1s-2p$                                    &1.473 &8.417 &SS~Cyg &648, 2307\\
Si XIII    &$1s^{2}~^{1}S-1s2p^{1}P$                  &1.865 &6.648 &U~Gem &3767\\
& & & &SS~Cyg& 648, 2307\\
Si XIV    &$1s-2p$                                    &2.007 &6.177 &U~Gem &647, 3767\\
& & & &SS~Cyg &648\\
Fe XXII  & K$_{\alpha}$                               &6.627 &1.871 &SS~Cyg &2307\\
\bottomrule 
\end{tabular}
\end{table*}

\subsection*{{RXTE} Spectral Analysis}
Almost all observations showed a  low count rate ($< 1$ cts/s) in the 
energy band $> 25$ keV  of the PCA spectra. The same, or no counts, were observed in the 40-150 keV of the
 HEXTE spectra. Therefore, except for Obs. ID 50012-01-01-00 of SS~Cyg, we performed the spectral analysis considering 
only the PCA spectra with energy range up to 25 keV. 

Because \textit{RXTE} Obs. ID 50012-01-01-00 of SS~Cyg is simultaneous to \textit{Chandra} 
Obs. ID 646, we have analysed the 0.4-8 keV HETG/ACIS \textit{Chandra} spectral band  
together with the 8-60 keV PCA and 60-150 keV HEXTE energy band of \textit{RXTE}.  


{
The spectrum of U~Gem Obs. ID 80011-01-02-00, SS~Aur Obs. ID 30026-03-01-00, and SS~Cyg 
Obs. ID 10040-01-01-000 were well described, in the $\sim$ 5--25 keV energy band, by only one 
 \textsc{compTT} component. On the other hand, the other three spectra of SS~Cyg 
(Obs. ID 50012-01-01-00, 10040-01-01-001, and 10040-01-01-00) required a second 
spectral component. For all these three observations, we fit a model consisting of (1) a blackbody 
(\textsc{bbody}) plus a Comptonization (\textsc{compTT}) component 
or (2) a sum of two Comptonization (\textsc{compTT}) components.

In general, taking into account the $\chi^2$-statistic and the range of physical parameters, the best fits were found for the second case -- that is, 
when two \textsc{compTT} components were used to describe the total spectra. Namely, the best-fits were found 
when both the electron temperature $kT_e$ and the optical depth $\tau$ of the \textsc{compTT} model were tied between the two components. 
{Figure \ref{fig:compCVs} shows the the geometry of our  Comptonization model.   
We definitely establish that there is a disk which supplies the seed photons for the transition layer (Compton cloud). While red lines indicate the trajectories of the photons which illuminate the transition layer coming from the white dwarf (WD) surface. As a result the emergent X-ray spectrum formed in the transition layer  as a result of up-scatering of the seed photons of the disk and WD surface off  hot electrons of the Compton cloud.  }

In order to assess the statistical significance of the second \textsc{compTT} component, we 
computed the probability of chance improvement of the $\chi^2$ by means of an F-test. It is worth stressing that the F-test to be used in this case is not the one that uses the $\Delta\chi^2$ as test statistics (let us call it \emph{independent} F-test, or I-F-test for short). Indeed, in order to correctly use the I-F-test, the $\chi^2$ variables must be linearly independent. In our case the two components share some parameters, therefore are not independent by construction, and therefore we must use another test statistics: the \emph{dependent} F-test (D-F-test for short).  This test statistics is defined in terms of the ratio between the normalized $\chi^2$ \citep[see, eg,][]{Barlow,NR}, and has been already successfully used for the assessment of the statistical significance of multiplicative components (that, too, are dependent components by construction) by \citet{Orlandini12} and \citet{Iyer15} among others.

\begin{figure}
\centering
\includegraphics[width=0.9\textwidth]{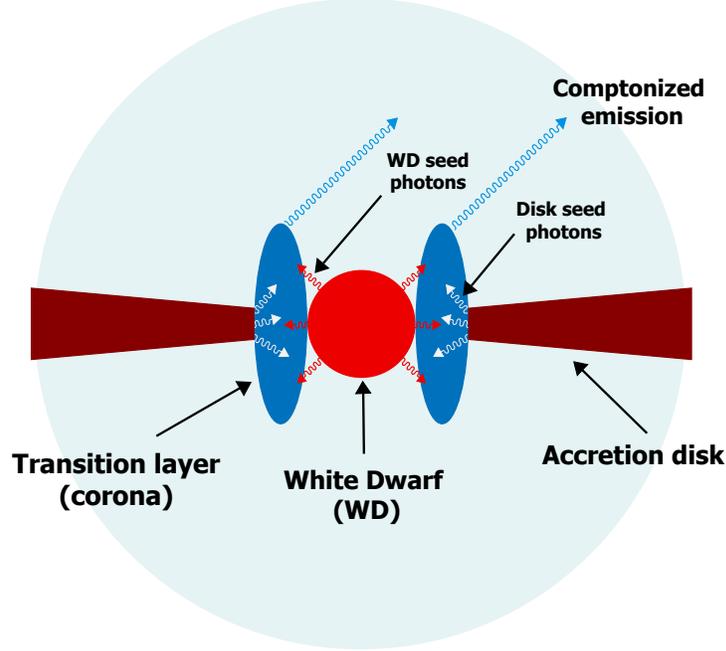}
\caption{Geometry of the Comptonization framework in nmCVs. Both the seed photons coming from the inner part of the accretion disk and the seed photons coming from (or close) the white dwarf surface are up-scattered by thermal electrons present in the transition layer (or corona), which is located between the WD surface and the accretion disk. The light blue sphere around the system represents the region wherein relatively cold material is outflowing, where lines are formed.} 
\label{fig:compCVs}
\end{figure}

The probability of chance improvement of the $\chi^2$ by adding the second \textsc{compTT}
component, evaluated by means of a D-F-test, is equal to 3$\times10^{-8}$ in  
SS~Cyg Obs 50012-01-01-00, 0.07 in Obs. 10040-01-01-001, and 0.02 
in Obs. 10040-01-01-00.

From these values we see that the second \textsc{compTT} component is statistically significant (equivalent to 5.4$\sigma$ for a one-tailed test) only for the 50012-01-01-00 SS~Cyg observation. This does not mean that this component is not present also in the other two \textit{RXTE} observations, but only that the SNR in those two latter observations does not allow its detection above a confidence level of about 2$\sigma$.
}

\begin{figure*}
\center
  \includegraphics[width=8.8cm]{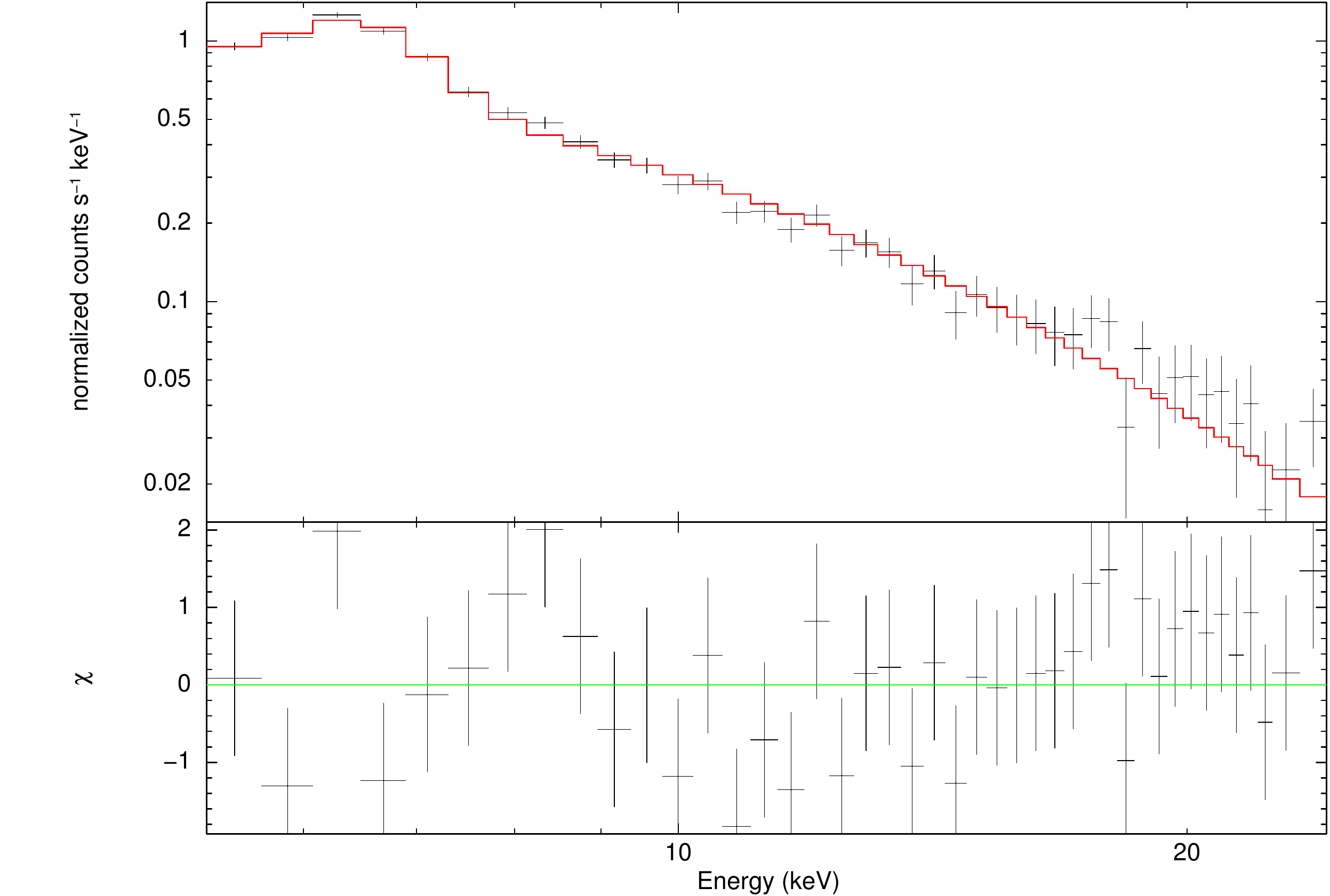}
   \includegraphics[width=8.8cm]{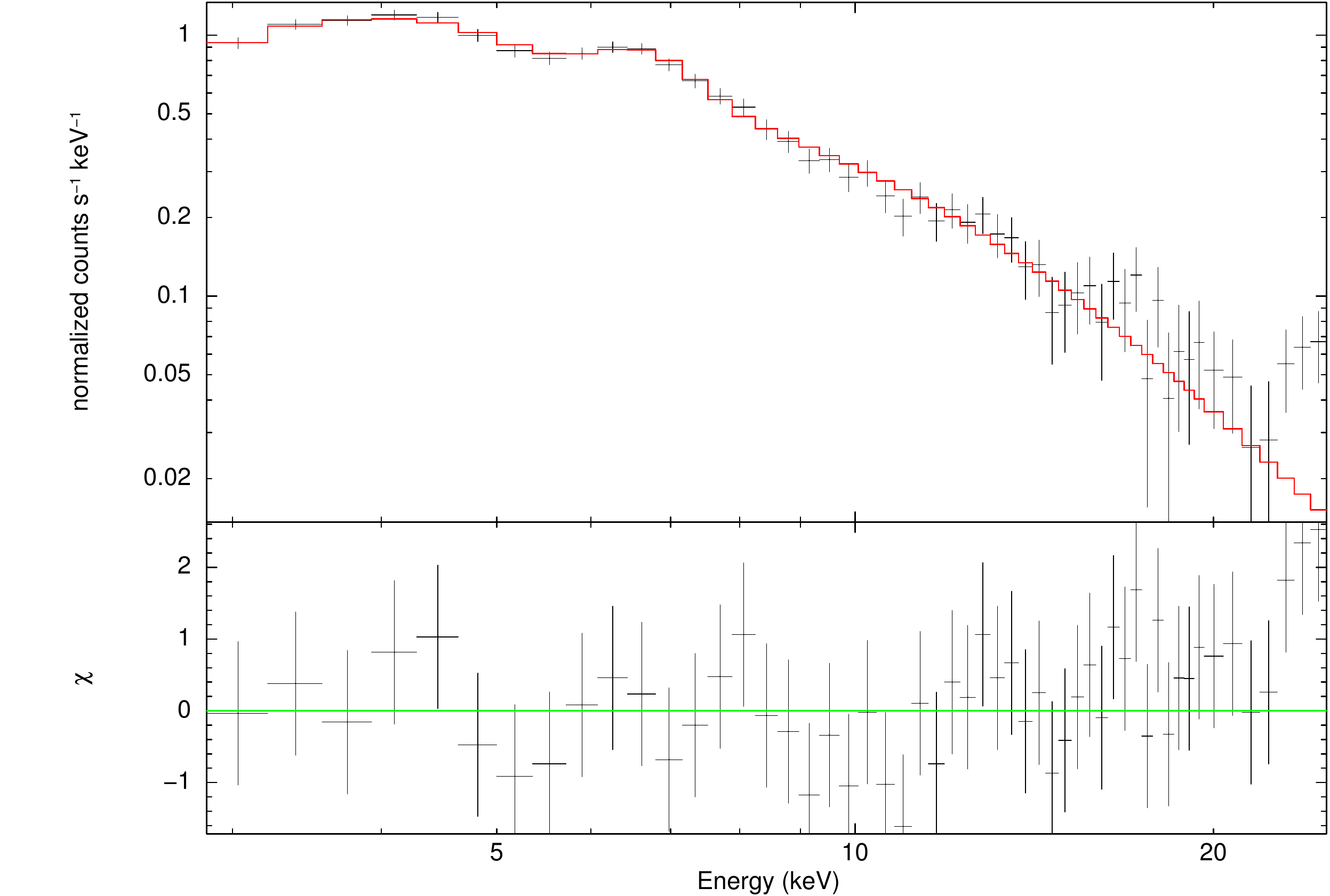}
    \caption{Spectral fits of the \textit{RXTE}/PCA data for U~Gem in outburst (\textit{left panel}) and SS~Aur in quiescence (\textit{right panel}). 
     The solid \textit{red} line shows the total \textsc{compTT+Gaussian} model (see Table \ref{tab:RXTEcontinuum} 
     and \ref{tab:RXTElines}). The lower 
     panels show the residuals of the data vs. model.}
 \label{fig:specUGemSS}  
\end{figure*}

\begin{figure*}
\center
    \includegraphics[width=8.8cm]{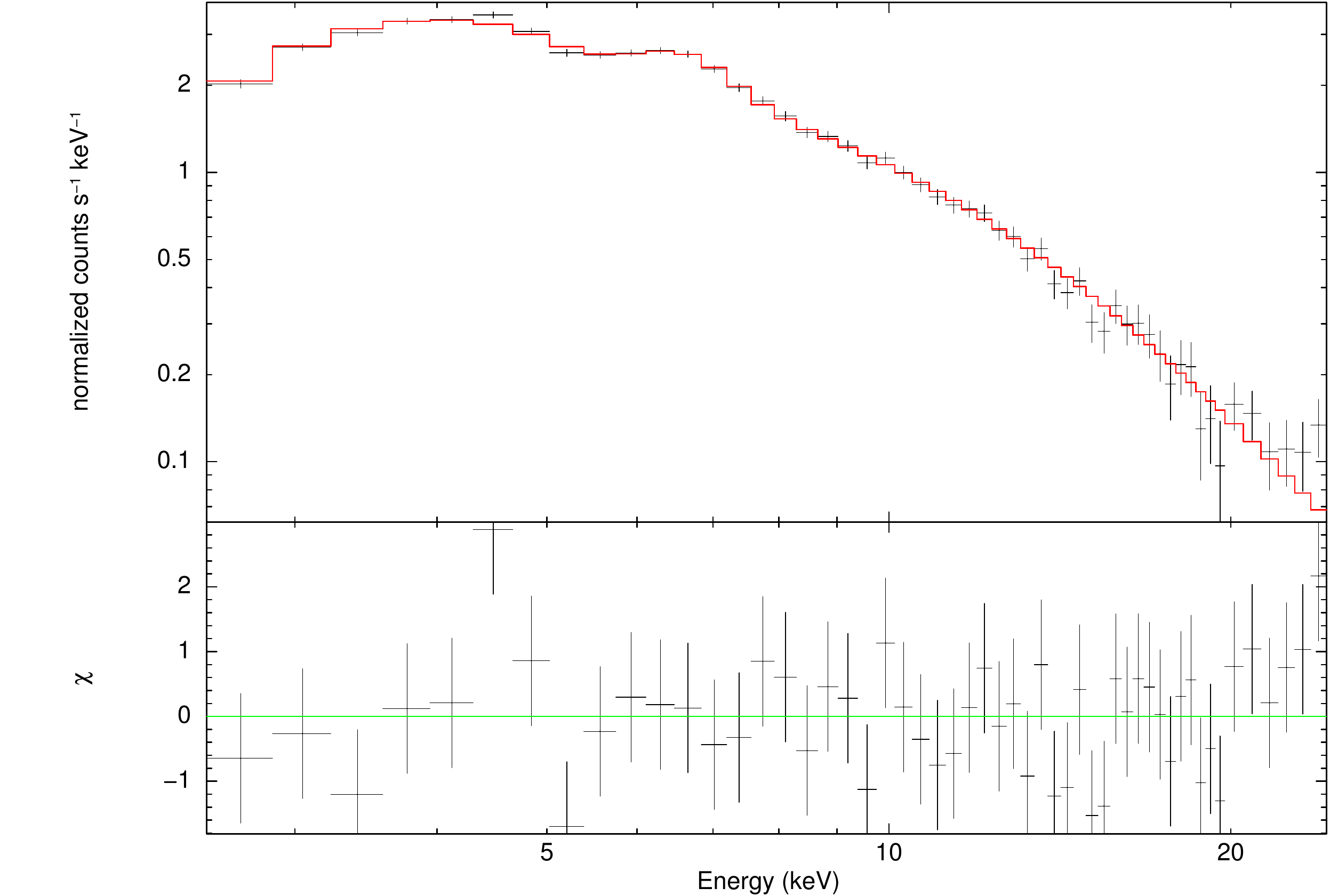}
\includegraphics[width=8.8cm]{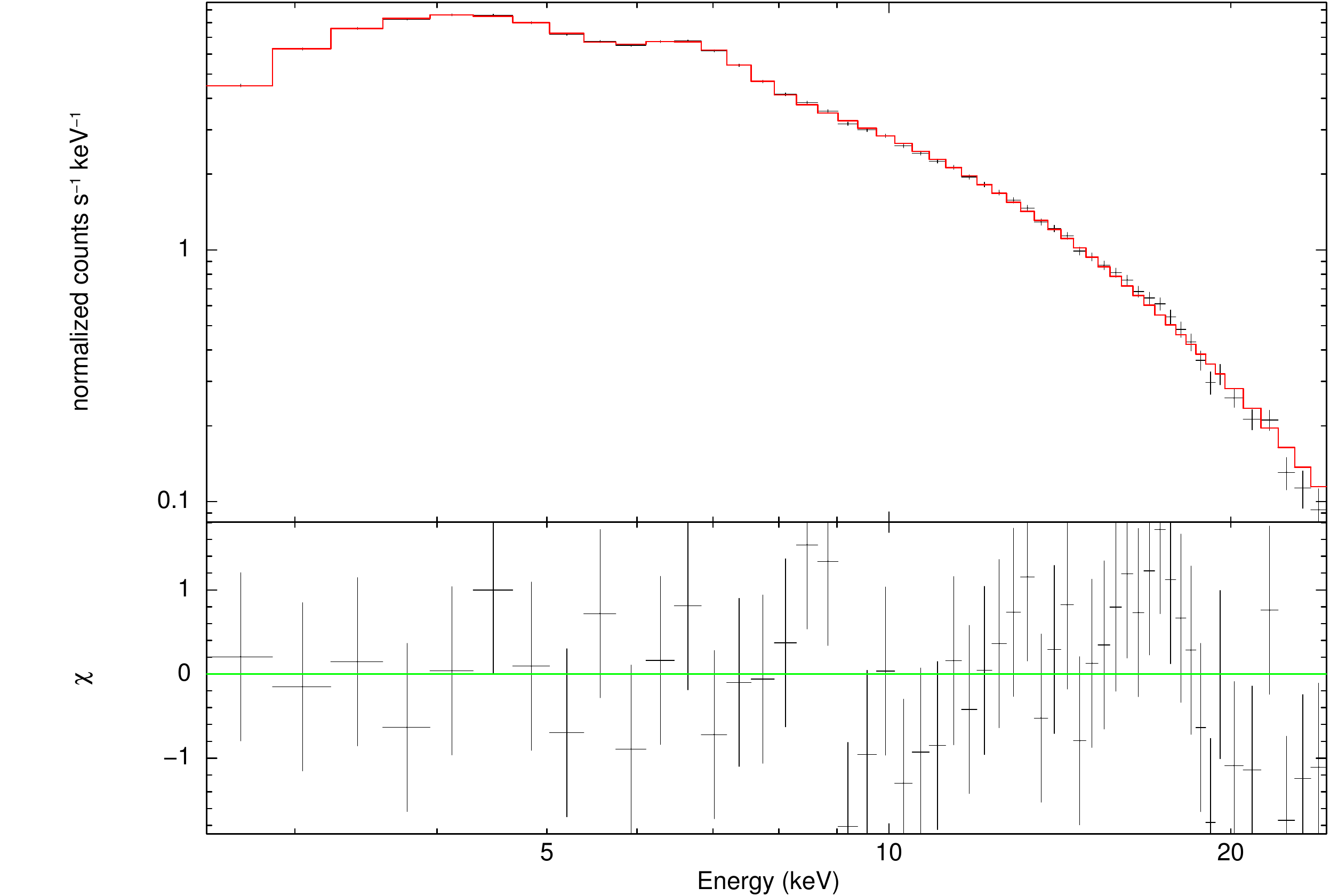}
 \includegraphics[width=8.8cm]{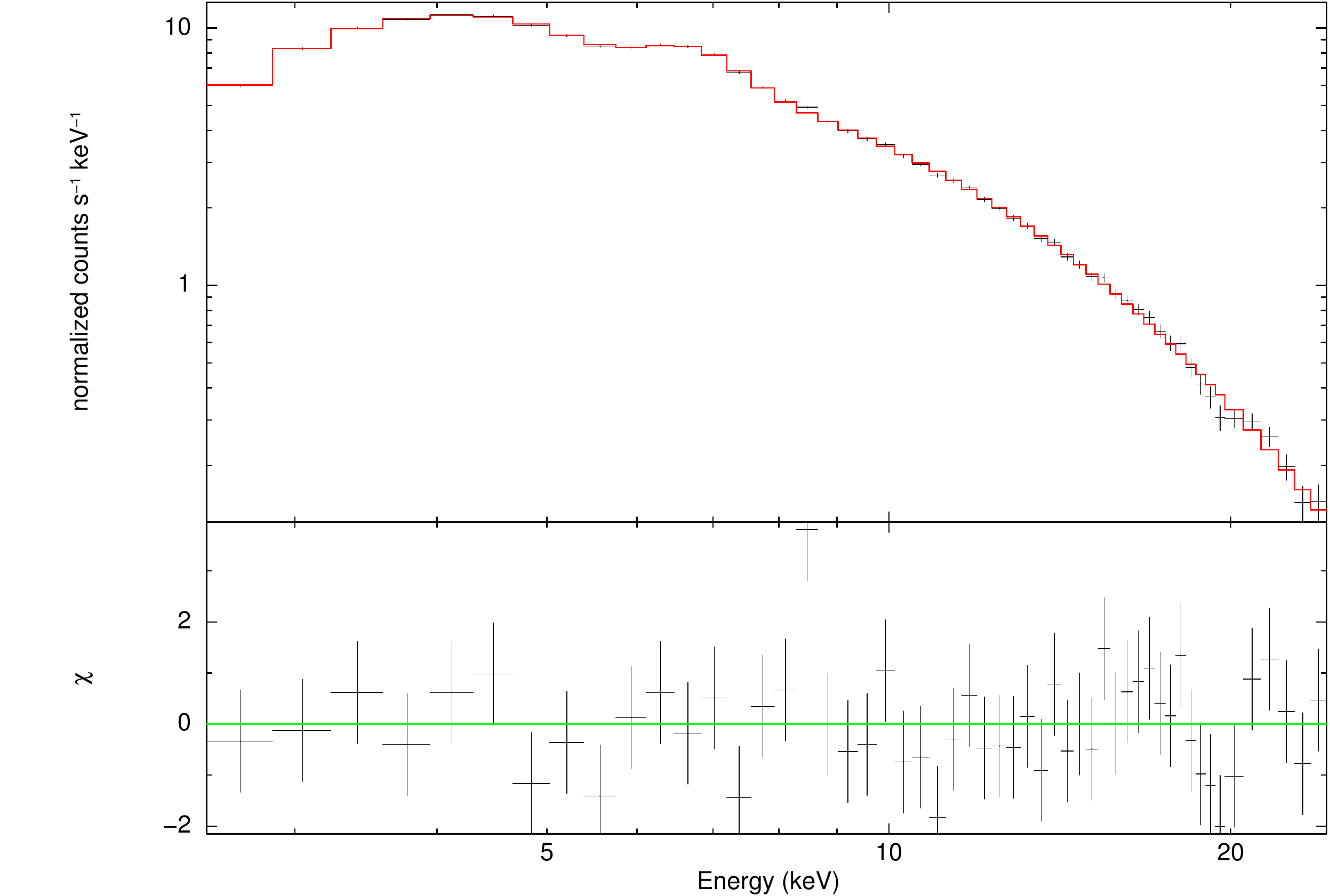}
    \caption{Spectral fits to the \textit{RXTE}/PCA spectra of SS~Cyg in outburst: Obs. ID 10040-01-01-000 
    (\textit{left top panel}), 10040-01-01-001 (\textit{right top panel}), and (10040-01-01-00
    \textit{lower panel}). The solid \textit{red} line shows the \textsc{compTT+Gaussian}  total model in Obs. ID 10040-01-01-000, 
      and \textsc{compTT+compTT+Gaussian} in Obs. ID 10040-01-01-001 and 10040-01-01-00 (see Table \ref{tab:RXTEcontinuum} 
     and \ref{tab:RXTElines}). The lower 
     panels show the residuals of the data vs. model.}
 \label{fig:specSSCyg10040}  
\end{figure*}

\begin{figure*}
\center
   \includegraphics[width=14cm]{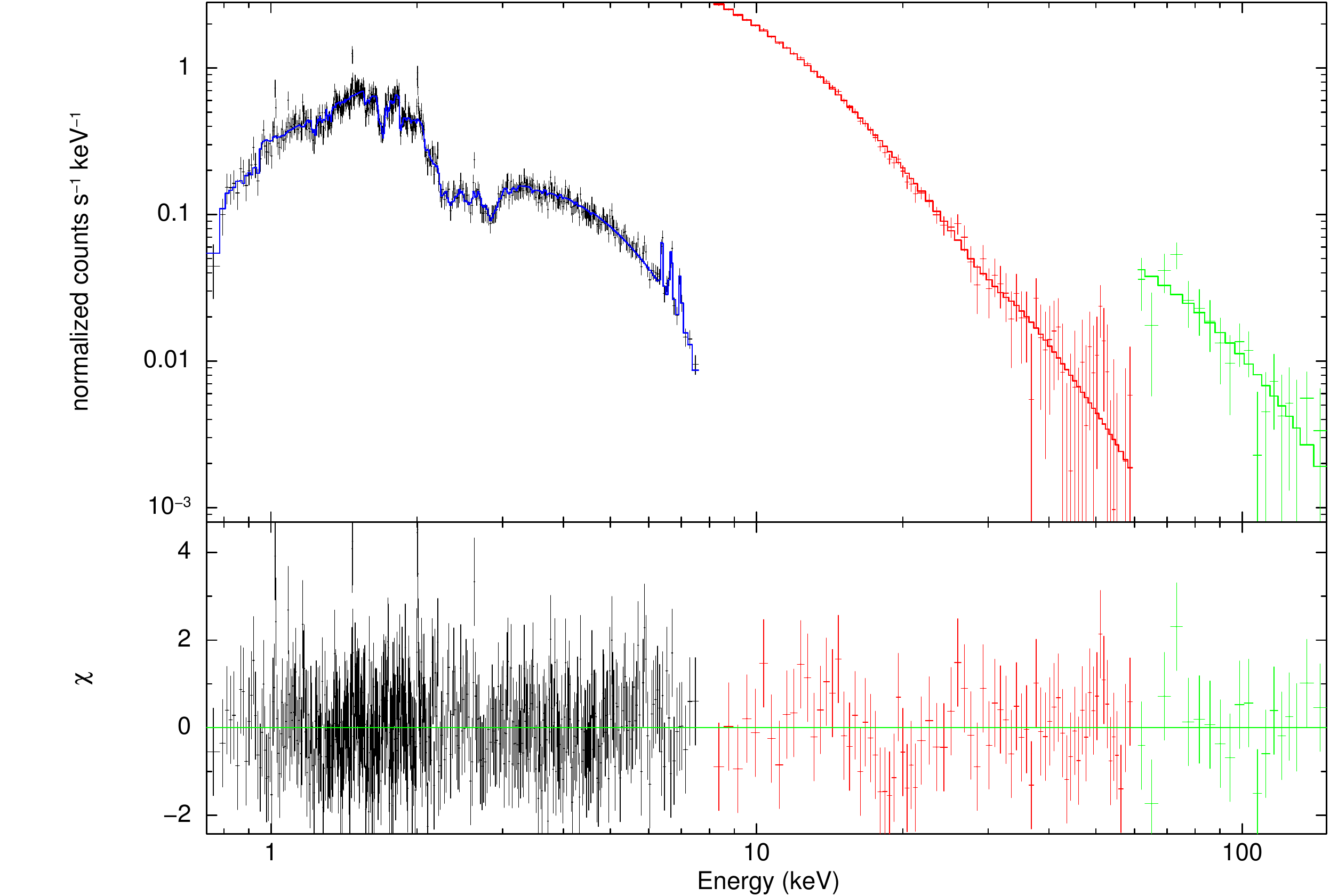}
     \caption{Spectral fit to the simultaneous \textit{Chandra}/\textit{RXTE} 
     (Obs. ID 646/Obs. ID 50012-01-01-00) spectra of SS~Cyg, using the  
     \textsc{compTT+compTT+gaussian+gaussian+gaussian} total model. The \textit{black}, \textit{red} and \textit{green} 
     points correspond to 0.4-8.0 keV HETG/ACIS, 8.0-60 keV PCA 
     and 60-150 keV HEXTE data, respectively. The \textit{blue}, \textit{red} and \textit{green} lines correspond to the best spectral fit (see column 6 of 
     Table \ref{tab:RXTEcontinuum} and \ref{tab:RXTElines}). The lower 
     panel show the residuals of the data vs. model}
 \label{fig:specSSCygSimul}  
\end{figure*}

\begin{figure*}
\center
   \includegraphics[width=18cm]{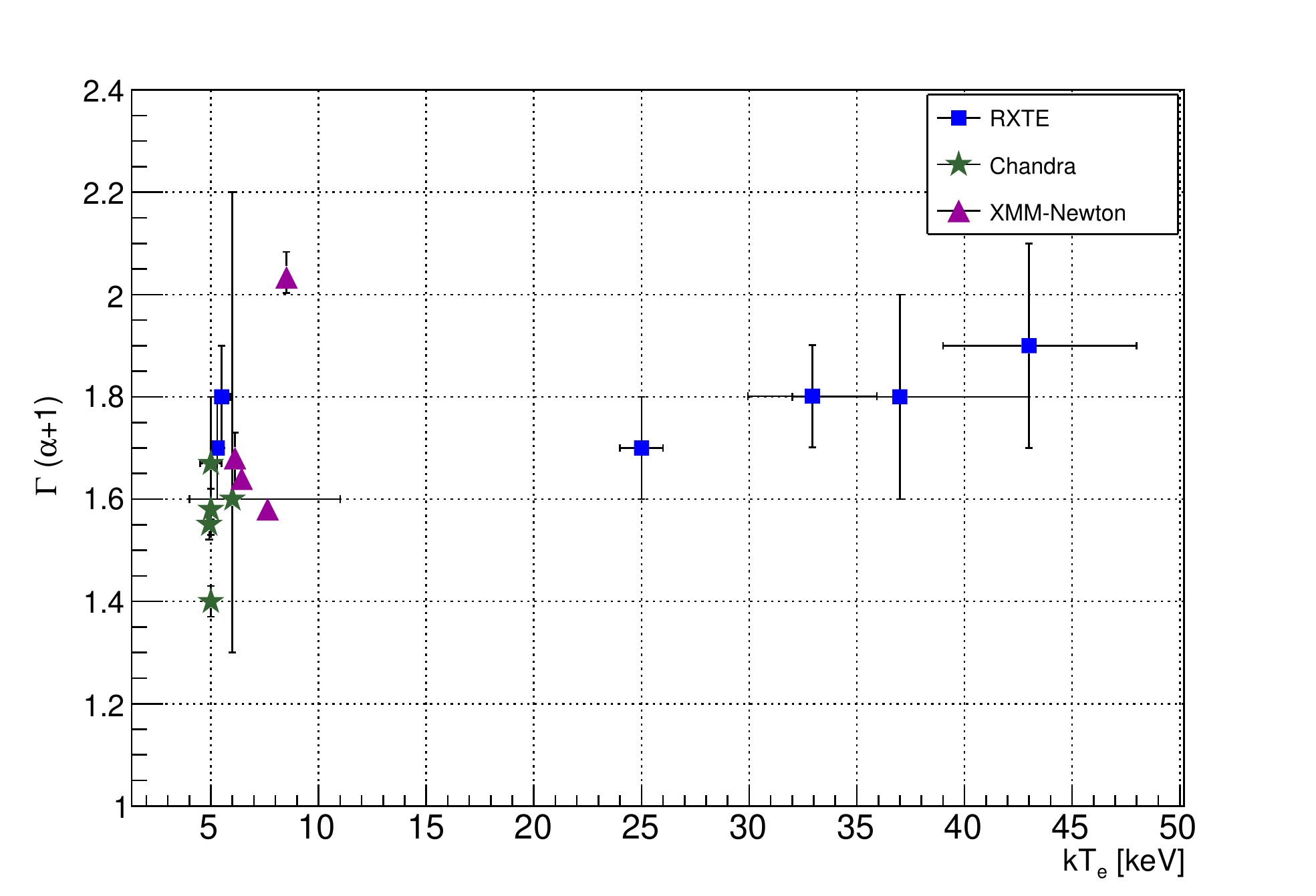}
    \caption{Photon indices $\Gamma$ vs. the best-fit electron temperatures kT$_e$ in the Compton cloud (Transition layer): 
    the blue, green and magenta points correspond to \textit{RXTE}, \textit{Chandra} and \textit{XMM-Newton} spectra analyzed, 
    respectively (see Tables \ref{tabT4:GanmCV}, \ref{tab:UGemChandraCOMPTT}, \ref{tabT4:SSCOMPTT} and \ref{tab:RXTEcontinuum}).}
 \label{fig:plotGammakTeRXTE}  
\end{figure*}

Table \ref{tab:RXTEcontinuum} shows the best-fit parameters and fit quality for each source and observation. 
Figure \ref{fig:specUGemSS} shows the spectral fits to U~Gem and SS~Aur. 
Figure \ref{fig:specSSCyg10040} shows the spectral fits to the three observations of SS Cyg in outburst. 
It is important to stress that these three observations are consecutive. 
Figure \ref{fig:specSSCygSimul} shows the spectral fit to the 0.4$-$150 keV simultaneous 
\textit{Chandra}/\textit{RXTE} spectra of SS Cyg in quiescence. 
Finally, Figure \ref{fig:plotGammakTeRXTE} shows the 
photon index $\Gamma$ as a function of the best-fit electron temperatures kT$_e$ of  the TL for all sources analyzed using our nmCV sample.  

Because of the low energy resolution of PCA, the emission iron lines expected in the $\sim$6.4$-$7.0 keV 
energy range are not well resolved. Therefore, only one Gaussian component is required to account 
for the excesses in this energy range. Table \ref{tab:RXTElines} shows the line energy found for 
each observation, wherein for the simultaneous \textit{Chandra}/\textit{RXTE} 
spectra we exceptionally show the three iron emission lines observed by HETG/ACIS \textit{Chandra}.   

\begin{sidewaystable}
\scriptsize
\caption{Spectral analysis of \textit{RXTE} observations using thermal \textsc{compTT} component(s).}
\label{tab:RXTEcontinuum}
\begin{tabular}{lllllllll}
\toprule
    &    &       &U~Gem                                     &SS~Aur                                          &SS~Cyg                                  &SS~Cyg                                            &SS~Cyg                                 SS~Cyg\\  
Obs. ID          &                     &             &80011-01-02-00                         &30026-03-01-00                            &50012-01-01-00                      &10040-01-01-000                              &10040-01-01-001                    &10040-01-01-00\\ 
State              &                     &             &outburst                                     &quiescence                                    &quiescence$^{*}$                    &outburst                                             &outburst &outburst\\ 
Component &Parameter &Unit & & & & & & \\
\midrule
CompTT$_1$ & kT$_{s1}$     &keV        &$0.23_{-0.03}^{+0.03}$           &$0.05_{-0.01}^{+0.01}$                &$0.11_{-a}^{+0.06}$                   &$0.20_{-0.01}^{+0.01}$                     &$0.31_{-0.05}^{+0.05}$           &0.29$_{-0.03}^{+0.03}$\\
\vspace{1.5mm}   
                & kT$_{e1}$            &keV        &$43_{-4}^{+5}$                        &37$_{-5}^{+6}$                             &32.8$_{-0.6}^{+0.5}$                  &$25_{-1}^{+1}$                                  &$5.3_{-0.3}^{+0.3}$                 &5.4$_{-0.3}^{+0.3}$\\ 
\vspace{1.5mm}
                & $\tau_1$                 &           &$0.83_{-0.16}^{+0.26}$            &1.0$_{-0.1}^{+0.1}$                     &1.21$_{-0.01}^{+0.01}$              &$1.82_{-0.06}^{+0.06}$                      &$4.8_{-0.2}^{+0.2}$                &4.5$_{-0.2}^{+0.2}$\\  
\vspace{1.5mm} 
                &norm$_1$            &              &$4_{-1}^{+1}\times10^{-4}$      &$8_{-1}^{+1}\times10^{-4}$         &$3.65_{-0.04}^{+0.04}\times10^{-3}$   &$1.02_{-0.03}^{+0.03}\times10^{-3}$  &$4_{-1}^{+18}\times10^{-3}$     &8.1$_{-0.7}^{+0.7}\times10^{-3}$\\             
 \vspace{1.5mm} 
 CompTT$_2$ & kT$_{s2}$    &keV      &                                                  &                                                  &0.89$_{-0.02}^{+0.02}$                 &                                                         &$0.65_{-0.03}^{+0.03}$           &0.76$_{-0.04}^{+0.04}$\\
\vspace{1.5mm}   
                & kT$_{e2}$          &keV        &                                                  &                                                   &= kT$_{e1}$                                  &                                                         &= kT$_{e1}$                             &= kT$_{e1}$\\ 
\vspace{1.5mm} 
                & $\tau_2$             &             &                                                  &                                                    &= $\tau_1$                                   &                                                         &= $\tau_1$                               &= $\tau_1$\\ 
\vspace{1.5mm}
                &norm$_2$           &              &                                                  &                                                    &5.0$_{-0.1}^{+0.1}\times10^{-4}$  &                                                        &$3_{-2}^{+3}\times10^{-3}$      &2$_{-1}^{+3}\times10^{-3}$\\           
 \vspace{1.5mm}   
 Fit quality  &$\chi^2$/d.o.f     &             &38/36                                     &40/53                                               &503/554                                     &42/46                                                   &42/44                               &50/44\\
                 &$\chi^2_{red}$     &             &1.06                                       &0.88                                                 &0.91                                            &0.91                                                     &0.94                                 &1.13\\ 
Energy range &                      & keV      &$5-25$                                   &$2.5-25$                                          &$0.4-150$                            &$2.5-25$                                              &$2.5-25$                          &$2.5-25$\\                                                
\bottomrule
\end{tabular}
\begin{itemize}
\item[ ] Uncertainties at 90\% confidence level. The respective emission lines present in the spectra are presented in Table \ref{tab:RXTElines}. 
\item[a] Parameter pegged at hard limit.
\item[*] Simultaneous RXTE/Chandra observation.  
\end{itemize}
\end{sidewaystable}


\begin{sidewaystable}
\centering
\caption{Spectral analysis of the \textit{RXTE} observations:  Gaussian component(s).}
\label{tab:RXTElines}
\footnotesize
\begin{tabular}{lllllllll}
\toprule
    &    &       &U~Gem                                     &SS~Aur                                        & SS Cyg                                      & SS Cyg                                              & SS Cyg                                &SS Cyg\\  
Obs. ID          &                     &             &80011-01-02-00                         &30026-03-01-00                           &50012-01-01-00                         &10040-01-01-000                                &10040-01-01-001                  &10040-01-01-00\\ 
State              &                     &             &outburst                                     &quiescence                                   &quiescence$^{*}$                   &outburst                                               &outburst                                &outburst\\ 
Component &Parameter &Unit & & & & & & \\
\midrule
Gaussian$_1$        & E$_{L1}$              &keV       &6.54$_{-0.06}^{+0.05}$           &6.63$_{-0.13}^{+0.14}$                            &$6.39_{-0.01}^{+0.01}$                 &$6.68_{-0.09}^{+0.09}$            &$6.64_{-0.03}^{+0.03}$           &6.64$_{-0.03}^{+0.03}$\\
\vspace{1.5mm}                                
                      & $\sigma_{L1}$       &keV       &0.36$_{-0.06}^{+0.05}$           &$0.32_{-{a}}^{+0.19}\times10^{-2}$   &$1.8_{-1.7}^{+1.1}\times10^{-2}$   &$0.38_{-0.12}^{+0.12}$            &$0.33_{-0.05}^{+0.05}$           &0.33$_{-0.04}^{+0.04}$\\
\vspace{1.5mm}
 Gaussian$_2$      & E$_{L2}$               &keV       & & &6.67$_{-0.02}^{+0.01}$ & & &\\ 
\vspace{1.5mm}                                
                     & $\sigma_{L2}$       &keV      & & &3.4$_{-1.1}^{+1.6}\times10^{-2}$ & & &\\
\vspace{1.5mm}
 Gaussian$_3$     & E$_{L3}$      &keV      & & &6.97$_{-0.02}^{+0.01}$ & & &\\  
\vspace{1.5mm}                               
                   & $\sigma_{L3}$        &keV      & & &2.12$_{-a}^{+0.02}\times10^{-2}$ & & &\\
\bottomrule 
\end{tabular}
\begin{itemize}
\item[ ] Uncertainties at 90\% confidence level.  
\item[a] Parameter pegged at hard limit.
\item[*] Simultaneous Chandra/RXTE (Obs. ID 646/50012-01-01-00) spectra. The lines present in the HETG/ACIS \textit{Chandra} spectrum are shown.  
\end{itemize}
\end{sidewaystable}


\begin{table*}
\centering\small

\caption{Best-fitting plasma temperature kT in the nmCVs of our sample obtained by thermal Comptonization (this paper) and some previous modeling (see  section \ref{sec:sources}).}
\label{tab:Tcomp}
\scriptsize
\begin{tabular}{llllll} 
\hline
\toprule 
Source  & State & kT$^{*}$ (keV) &Spectral Model &Observatory & Reference\\
\midrule
U~Gem &q$^{\dag}$ &$20$ &\textsc{mkcflow} &Chandra/HETG &\citet{Mukai2003} (Obs. ID 647)\\
              &q & $55^{+10}_{-10}$ &\textsc{mkcflow} &XMM-Newton/EPIC &\citet{pandel2005}\\
              &q &$15-38$ &\textsc{cemekl}, \textsc{mkcflow} &XMM-Newton/EPIC, &\citet{Guver2006} (Obs. ID 0110070401, 647)\\
              & & & &Chandra/HETG &\\
              &o$^{\ddag}$&$5.4-29$ &\textsc{cemekl} &Chandra/HETG &\citet{Guver2006} (Obs. ID 3767)\\                       
              &q &$25.82^{+1.98}_{-1.43}$ &\textsc{mkcflow} &XMM-Newton/EPIC &\citet{byckling2010}\\
              &q &$0.78^{+0.03}_{-0.01}$ &\textsc{mekal} &XMM-Newton/EPIC &\citet{byckling2010}\\
              &q &$16.5^{+4.49}_{-3.31}$ &\textsc{apec} &SUZAKU/XIS &\citet{Xu2016}\\                            
              &q &$6.43^{+0.06}_{-0.06}$ &\textsc{compTT} &XMM-Newton/EPIC &this paper (Obs. ID 0110070401)\\ 
              &q &$5.0^{+0.5}_{-0.5}$ &\textsc{compTT} &Chandra/HETG &this paper (Obs. ID 647)\\  
              &o &$6^{+5}_{-2}$ &\textsc{compTT} &Chandra/HETG &this paper (Obs. ID 3767)\\ 
              &o &$43^{+5}_{-4}$ &\textsc{compTT} &RXTE/PCA &this paper (Obs. ID 80011-01-02-00)\\                                             
SS~Cyg &q &$80$ &\textsc{mkcflow} &Chandra/HETG &\citet{Mukai2003} (Obs. ID 646)\\
              &q &$20.4_{-2.6(stat)+3.0(sys)}^{+4.0(stat)+3.0(sys)}$ &\textsc{cvmekal} &SUZAKU/XIS  &\citet{ishida2009}\\
              &o &$6.0^{+1.3}_{-0.2}$ &\textsc{cvmekal} &SUZAKU/XIS &\citet{ishida2009}\\              
              &q &10.44$^{+0.16}_{-0.17}$ &\textsc{mekal} &SUXAKU/XIS &\citet{byckling2010}\\
              &q &41.99$^{+1.20}_{-0.76}$ &\textsc{mkcflow} &SUXAKU/XIS &\citet{byckling2010}\\
              &o &8.15$^{+1.23}_{-0.91}$ &\textsc{apec} &SUZAKU/XIS &\citet{Xu2016}\\ 
              &q &$7.64^{+0.05}_{-0.05}$ &\textsc{compTT} &XMM-Newton/EPIC &this paper (Obs. ID 0111310201)\\ 
              &q &5$^{**}$ &\textsc{compTT} &Chandra/HETG &this paper (Obs. ID 646)\\ 
              &o &5$^{**}$ &\textsc{compTT} &Chandra/HETG &this paper (Obs. ID 648)\\ 
              &o &5$^{**}$ &\textsc{compTT} &Chandra/HETG &this paper (Obs. ID 2307)\\ 
              &o &25$^{+1}_{-1}$ &\textsc{compTT} &RXTE/PCA &this paper (Obs. ID 10040-01-01-000)\\   
              &o &5.3$^{+0.3}_{-0.3}$ &\textsc{compTT} &RXTE/PCA &this paper  (Obs. ID 10040-01-01-001)\\   
              &o &$5.4^{+0.3}_{-0.3}$ &\textsc{compTT} &RXTE/PCA &this paper (Obs. ID 10040-01-01-00) \\   
              &q &$32.8^{+0.5}_{-0.6}$ &\textsc{compTT} &Chandra/RXTE$^{\S}$ & this paper$^{\S}$\\                                                                            
VW~Hyi &q &$>5$ &Bremsstrahlung &EXOSAT &\citet{Woerd1987}\\
              &q &$2.17^{+0.15}_{-0.15}$ &Raymond-Smith &ROSAT &\citet{Belloni1991}\\
              &? &$4-10$ &two- \textsc{mekal} &ASCA &\citet{Hasenkopf2002}\\              
              & & & or Raymond-Smith & &\\
              &q &$6-8$ &\textsc{cemekl, cevmkl} &XMM-Newton/EPIC &\citet{Pandel2003} (Obs. ID 0111970301)\\
              & & & and \textsc{mkcflow} & &\\
              &q & $8.2^{+0.3}_{-0.3}$ &\textsc{mkcflow} &XMM-Newton/EPIC &\citet{pandel2005} (Obs. ID 0111970301)\\
              &q &$5.79^{+4.71}_{-2.14}$ &\textsc{apec} &SUZAKU/XIS &\citet{Xu2016}\\              
              &q &$8.58^{+0.14}_{-0.14}$ &\textsc{compTT} &XMM-Newton/EPIC &this paper (Obs. ID 0111970301)\\
              &q & 5-9 &\textsc{cevmkl,vmcflow} &XMM-Newton/EPIC &\citet{Nakaniwa2019} (Obs. ID 0111970301)\\
              & & & &SUZAKU/XIS &\\ 
SS~Aur &q &6.35$^{+0.40}_{-0.40}$ &\textsc{mekal} &SUZAKU/XIS &\citet{byckling2010}\\
              &q &$23.47_{-3.02}^{+4.01}$ &\textsc{mkcflow} &SUZAKU/XIS &\citet{byckling2010}\\
              &q &$8.48^{+5.34}_{-2.61}$ &\textsc{apec} &SUXAKU/XIS &\citet{Xu2016}\\
              &q &$6.12^{+0.13}_{-0.13}$ &\textsc{compTT} &XMM-Newton/EPIC &this paper (Obs. ID 0502640201)\\ 
              &q &$37^{+6}_{-5}$ &\textsc{compTT} &RXTE/PCA &this paper (Obs. ID 30026-03-01-00)\\                                          
\bottomrule 
\hline
\end{tabular} 
\begin{itemize}
\item[*] For multi-temperature plasma models, the temperature kT correspond to the maximal temperature ($kT_{\rm{max}}$)
to that the plasma is heated up. 
\item[**] Parameter frozen.
\item[$\dag$] Source observed in quiescence state.
\item[$\ddag$] Source observed in outburst state.
\item[$\S$] Simultaneous Chandra HETG/ACIS Obs. ID 646  and {\it RXTE} (PCA and HEXTE) Obs. ID 50012-01-01-00. 

\end{itemize} 
\end{table*}

\section{The transition layer in a WD and spectral index of the expected emergent spectrum in the Comptonization scenario}
\label{sec:model}
As \citet{ft11}, hereafter FT11, pointed out  the energy release in the {TL} of a NS determines the spectral index of the emergent spectrum.  In similar way we  use the TL model to estimate the index for a case of a WD. But in this case instead of the reflection inner boundary of the TL at  a NS  we use the pure absorption boundary at a WD.  
 FT11 
demonstrated that the energy flux  per unit surface area of the TL (corona) can be found as 
\begin{equation}
Q_{cor}=20.2\int_0^{\tau_0}\varepsilon(\tau)T_e(\tau)d\tau,
\label{energy_release_ in_TL}
\end{equation}
 where   $T_e(\tau)$, $\varepsilon(\tau)$ and $\tau_0$  are the plasma (electron) temperature, the radiation density  distributions in the TL and its Thomson optical depth, respectively.

We obtain the energy distribution $\varepsilon(\tau)$ as a solution of the diffusion equation 
\begin{equation}
\frac{d^2\varepsilon}{d\tau^2}=-\frac{3Q_{cor}}{c\tau_0}
\label{energy_distribution_ in_TL}
\end{equation}
where $c$ is the speed of light.

We should also add the two boundary conditions at the inner TL boundary, which can be an absorbed surface at  WD $\tau=\tau_0$ and the outer boundary $\tau=0$.  They are correspondingly:
\begin{equation}
\frac{d\varepsilon}{d\tau}+\frac{3}{2}\varepsilon|_{\tau=\tau_0}=0, 
\label{inner_bound_cond}
\end{equation}
\begin{equation}
\frac{d\varepsilon}{d\tau}-\frac{3}{2}\varepsilon |_{\tau=0}=0. 
\label{outer_bound_cond}
\end{equation}
We find  the solution 
$\varepsilon(\tau)$ of equations (\ref{energy_distribution_ in_TL}-\ref{outer_bound_cond})

\begin{equation}
\varepsilon(\tau)= \frac{Q_{cor}}{c}\{1+ 3/2~\tau_0[(\tau/\tau_0)-(\tau/\tau_0)^2]\}~~~
\label{mod_epsilon_tau_1}
\end{equation}

Thus integration of $\varepsilon(\tau)$  gives us
\begin{equation}
\int_0^{\tau_0}\varepsilon(\tau)d\tau= \frac{Q_{cor}}{c}\tau_0 (1+\tau_0/4). 
\label{int_epsilon}
\end{equation}

Now we can estimate the Comptonization Y-parameter in the TL using
 Eqs. (\ref{energy_release_ in_TL}) and  (\ref{int_epsilon}). 
We rewrite Eq. (\ref{energy_release_ in_TL}) using the mean value theorem as
\begin{equation}
Q_{cor}=20.2\hat{T}_e \int_0^{\tau_0}\varepsilon(\tau)d\tau
\label{energy_release_ in_TL_m}
\end{equation}
where $\hat{T}_e$ is the mean electron temperature in the TL.

Substitution of formula (\ref{int_epsilon}) in Eq. (\ref{energy_release_ in_TL_m}) 
leads to the following estimate  (Y-parameter)  in the TL [see a definition of  Y-parameter in 
\cite{rl79}]
\begin{equation}
\frac{k\hat{T}_e\tau_0(\tau_{0}/4+1)}{m_ec^2}=0.25
\label{y_parameter_estimate2}
\end{equation}


As \textbf{} we can see in Figure \ref{fig:plotGammakTeRXTE}, the photon index, $\Gamma$ 
shows only small 
deviations from 1.8: namely, for most of 
our sources $\Gamma= 1.8\pm 0.1$,
while the Compton cloud electron temperature varies from 5 to 45 keV.
In some sense this index behavior is similar to that observed in NSs  \citep{Titarchuk2014}. 

As already pointed out in classical work \citep{ft11},
spectral formation in plasma clouds of finite size (bounded medium)  is related to the distribution law of the number of scatterings that seed photons experience before  their escape. 
If $u_{av}$  denotes the average number of photon scatterings and the dimensionless scattering number is $u=N_e \sigma_T c t$,  then the distribution law for  $u\gg u_{av}$ is given by \citep{st80, st85}

\begin{equation}
 P(u)=A(u,\tau_0) e^{-\beta u}.
\label{prob_law}
\end{equation}
For  a diffusion regime when $\tau_0 > 1$,   the corresponding  $\beta=\lambda^2_1/3$, where $\lambda_1$ is the first eigenvalue of the diffusion space operator.
As reported in  \citet{st85}, the eigenvalue problem for photon diffusion in a slab
with the total optical depth, $\tau_0$ is derived from a solution of the differential equation 
for the zero-moment intensity

\begin{equation}
\frac{d^2J}{d\tau^2}+\lambda^2 J=0,
\label{eigenval_eq}
\end{equation}
with  boundary conditions $dJ/d\tau-(3/2)J=0$ and $dJ/d\tau+(3/2)J=0$,
for $\tau=0$ and $\tau=\tau_0$, respectively.
This leads to the trascendental equation for the eigenvalue $\lambda_n$, $n=1,2, 3$\ldots

\begin{equation}
 \rm{\tan}(\lambda_n\tau_0/2)=\frac{2}{3\lambda_n},
\end{equation}
which has the solution 
for $n=1$

\begin{equation}
\lambda_1= \frac{\pi}{2(\tau_0/2 + 2/3)}.
\label{eigen_val}
\end{equation}

But the spectral index $\alpha$ (photon index $\Gamma=\alpha+1$)  is 
\begin{equation}
\alpha=-3/2 + (9/4 +\beta/\theta)^{1/2}
\label{index_alpha}
\end{equation}
where $\beta=\lambda^2_1/3$  and $\theta=kT_e/(m_ec^2)$  \citep{st80}. 
Thus  
\begin{equation}
\alpha=-3/2 + (9/4 +\pi^2/[12(\tau_0/2 + 2/3)^2\theta]^{1/2}
\label{index_alpha_1}
\end{equation}
and using Equation (\ref{y_parameter_estimate2}) we obtain that $\alpha\lax 0.85$ (or $\Gamma \lax 1.85$).
\textbf{} This is precisely what we observed (see Fig. \ref{fig:plotGammakTeRXTE}).

\section{Discussion}  
\label{sec:discussion}

As previously mentioned, the Comptonization  model is not the standard radiative one  
currently used for the description of the continuum in nmCVs.
On the other hand, the Comptonization model is the standard one for fitting the  LMXBs spectra.

Considering that nmCVs share similarities with LMXBs -- that is, 
they have an accretion disk, a compact object  and a TL (corona) (which presence 
is evident from observation of non-eclipsed UV emission lines in eclipsing systems 
\citep[see, e.g.,][]{Warner1995,Mauche2000}) --  we look for spectral similarities 
between these two types of X-ray binaries. That is, our main goal is to seek a 
common physical process able to describe the continuum of both nmCVs and LMXBs.

Figure \ref{fig:compCVs} represents the Comptonization framework in nmCVs, in which the emission lines are produced far from the most central part of the system where the X-ray continuum is produced.

Because of the broad spectral X-ray energy band of PCA and HEXTE instruments, 
we have also analyzed some of \textit{RXTE} observations of our source sample  which is available in the archive
(see Table \ref{tab:obs}), including a simultaneous 0.4$-$150 keV \textit{Chandra}/\textit{RXTE} spectra of SS Cyg 
in quiescence state. The broadband, including harder ($> 15$ keV) X-rays, can provide a better 
understanding  of this scenario and physical parameters in terms of the Comptonization framework in these sources
-- mainly the electron temperatures and spectral indices.

All sources in our sample were observed in the quiescence state by \textit{XMM-Newton} Epic-pn. 
We found that only one thermal Comptonization component plus Gaussian components successfully 
fit the \textit{XMM-Newton} Epic pn spectra of two nmCVs: SS Cyg and SS Aur. 
{In terms of $\chi^2$-statistics, our model did not provide a perfect description 
of the VW~Hyi and U~Gem \textit{XMM-Newton} total spectra. In these two cases, the value of 
$\chi^2$ exceeds the critical value at 0.01 level of significance. However, the spectral analyses 
of U~Gem using \textit{Chandra} and \textit{RXTE} data show a satisfactory fit 
(see Table \ref{tab:UGemChandraCOMPTT} and \ref{tab:RXTEcontinuum} column 4). 
It is likely that the presence of several emission lines, besides the iron complex,
in the \textit{XMM-Newton} spectra of VW~Hyi and U~Gem worsened the fit quality.} 

It is important  to emphasize  that all  \textit{XMM-Newton} spectral fits show 
best-fit parameters with a mean seed photon temperature 
$<kT_s>$ of {$0.14 \pm 0.01$} keV, a mean optical depth $<\tau>$ of {$4.1 \pm 0.5$}, and a mean electron temperature $<kT_e>$ of {$7.2 \pm 0.6$} keV. The spectral fits {performed in the 1.5 to 15 keV energy range} of the \textit{XMM-Newton} spectra does not allow accurately  estimate the temperature of the seed photons.
In this case, the parameters appeared with a mean value $<kT_s>$ of $0.25 \pm 0.05$ keV, 
$<\tau>$ of $4.5 \pm 0.5$, and $<kT_e>$ of $5.9 \pm 0.5$ keV. Therefore, we obtained 
a better description of the seed photon temperature when considering the total Epic pn spectral 
band in our analyses.

The \textit{Chandra} HETG/ACIS spectra show 
parameters similar to those found by \textit{XMM-Newton}, with a mean $<kT_s>$ of 
$0.20 \pm 0.12$ keV, a mean optical depth $<\tau>$ of $5.2 \pm 0.2$, and a mean electron temperature $<kT_e>$ of $5.8 \pm 0.4$ keV. 
SS~Cyg was observed once in the quiescence state and twice in the outburst state by \textit{Chandra}. 
We did not obtain a perfect fit for the two observations in outburst (see Table \ref{tabT4:SSCOMPTT}).  
In addition, we did not observe a difference on the spectral continuum parameters between the two states. 
This can be due to the \textit{Chandra} effective energy band ($<$8 keV).   
Namely, we found that  the electron temperature  in the \textit{Chandra} spectra 
are not very well constrained by these fits. 

In the analyses of the \textit{RXTE} spectra, a second Comptonization component was necessary 
for the description of the total spectrum in three observation of SS~Cyg. In these cases, a satisfactory 
spectral fit was found if, both  $kTe$  and the optical depth $\tau$
 of the \textsc{compTT} model were tied between the two components. Namely, it means 
that the total $\sim$ 0.4$-$150 keV or 2.5$-$25 keV spectra of all  analysed nmCVs are characterized by only one spectral index $\alpha$ ($\alpha=\Gamma-1$). 

The \textit{RXTE} spectra of our sample showed a wide range of the electron temperatures, $kT_e$ laying in the 5$-$48 keV range. When one compares  values of $\alpha$ (or $\Gamma$)  between the different sources (see Table \ref{tab:RXTEcontinuum}, Fig. \ref{fig:plotGammakTeRXTE}), he/she does not find 
any correlation of $kT_e$ with $\Gamma$. 
It is also  worth noting that SS~Cyg shows $kT_e$ $\sim$ 
5 keV in outburst, U~Gem shows kT$_e$ of $43^{+5}_{-4}$ keV also in outburst state, and SS~Aur shows a 
temperature of $37^{+6}_{-5}$ keV in the quiescence state. Therefore, we establish that we do not find any correlation of the kT$_e$ with the spectral state using the data of our set of CVs. For sake of comparison,  we present in Table \ref{tab:Tcomp} the best-fitting plasma temperature found by our analyses together with the ones reported by some previous spectral modeling, for all CVs in our sample (see section \ref{sec:sources}). As shown in Table \ref{tab:Tcomp}, the spectral modeling present in the literature is not homogeneous and no clear correlation of the plasma temperature with spectral stage is found between the sources in previous spectral modeling as well.

In the \textit{RXTE} spectra, the first Comptonization component shows a mean value of the seed photon 
temperature kT$_{s1}$ of $0.20 \pm 0.04$ keV. This temperature  appears higher in the observations in outburst. 
SS Cyg shows kT$_{s1} \sim$ 2$-$3 times higher when in outburst. It is important to point out, however, that the \textit{RXTE} data do not precisely determine the seed photon temperatures if their values are much less than 1 keV because the lower limit of the \textit{RXTE} data is $\sim$ 3 keV. Therefore, all values of the seed photons kT$_s$ obtained through \textit{RXTE} data (see Table \ref{tab:RXTEcontinuum}) are upper limits.

In the three \textit{RXTE} observations of SS Cyg for which the second Comptonization component was required, the 
seed photon temperature kT$_{s2}$ of the second component appears with a mean value of $0.77 \pm 0.07$ keV. 
This component is not observed at the very early stage of the optical outburst, but reappears in the 
consecutive observations increasing its value to  0.76$_{-0.04}^{+0.04}$ keV in the last observation, 
closer to the outburst optical peak. In the observation of  the quiescence state, kT$_{s2}$ is equal to 
0.89$_{-0.02}^{+0.02}$ keV. We, therefore, did not observe a huge difference in the temperature kT$_{s2}$ 
between the different states. Interestingly, the \textit{Chandra} spectrum 
of U~Gem Obs. ID 3767 in outburst (see Table \ref{tab:UGemChandraCOMPTT}) shows a single 
Comptonization component with the seed photon temperature of $0.66^{+0.12}_{-0.10}$ keV. A transient hard component was reported in U~Gem during its 2004 outburst (see section \ref{sec:UGem} and  \citep{Guver2006}) and associated with the outburst state. We, on the other hand, observed this hard component also during the quiescence state in SS~Cyg. This is an evidence that this hard X-ray component is not a spectral feature of outburst states only.

In the Comptonization framework, the seed photons are up-scattered (Comptonized) by hot 
electrons of  a Compton cloud around the compact object (ST80). 
Our analyses showed that there are up to two seed photon components in nmCVs  presumably  coming from the 
internal and outer parts of the TL.
The corresponding  components are the results of the Comptonization of these soft photons{, characterized by their color temperature kT$_{s1}$ and kT$_{s2}$,} in the TL
located between the  WD surface and the inner part of the accretion disk. The electrons in the TL are characterized by a single electron  temperature 
as clearly showed by our analyses of the \textit{RXTE} spectra. 

The source of hard X-rays in quiescence is compact, as observed by X-ray light curves of 
eclipsing DNe \citep[see, e.g.][and references therein]{Mukai2017,Lewin2006}. Therefore, the 
Comptonized cloud (TL) should be compact.
In our interpretation, the seed photon component showing lowest temperatures (kT$_{s1} \sim$ 0.1$-$0.2 keV) 
is coming from the inner part of the accretion disk, while the second and transient component related to kT$_{s2}$
is coming from a more internal part of the system, closer or from the WD surface. 

 We tested, for the \textit{XMM-Newton} and \textit{RXTE} spectra, if an addition 
of an interstellar absorption component (\textsc{tbabs} model in XSPEC) would result in  
lower temperatures. We freeze the interstellar absorption parameter 
(NH) to the value expected in the line of sight of the source. In the \textit{XMM-Newton} spectra, 
we easily obtained satisfactory fits for VW~Hyi, U~Gem and SS~Aur sources.  In these cases, 
$kT_{s1}$ assumed values of $\sim$ 0.01 keV, but in addition to the observed increase of 
$\chi^2_{red}$, this parameter was not well constrained by the fit. In the \textit{RXTE} spectra, 
the inclusion of the \textsc{tbabs} component did not change the best-fit parameters.  
We obtained the same result when setting NH as a free parameter: in some of the fits it  assumed at a very low value of 10$^{19}$ atoms cm$^{-2}$, 
and did not affect the best-fit parameters in all fits.

Figure \ref{fig:specSSCygSimul} shows the total 0.4$-$150 keV spectrum of SS~Cyg 
in quiescence.  This spectrum corresponds to the simultaneous observations of \textit{Chandra} 
HETG/ACIS Obs. ID 646 and \textit{RXTE} (PCA and HEXTE) Obs. ID 50012-01-01-00. It is important to 
point out that the spectral analysis considering the simultaneous hard X-ray spectral band ($> 15$ keV) of \textit{RXTE} is superior and of great importance to constrain  the physical parameters -- e.g., the temperature of the electrons in the TL (kT$_e = 33^{+3}_{-3}$ keV) and the spectral index ($\Gamma= 1.8^{+0.1}_{-0.1}$), since considering only the HETG/ACIS spectrum it leads to lower values of kT$_e$ and $\Gamma$ (kT$_e = 5$ keV and $\Gamma = 1.55^{+0.02}_{-0.03}$); whereas the soft spectral band of \textit{Chandra} allows a better determination of the seed photons temperature kT$_{s1}$ (see Table \ref{tabT4:SSCOMPTT} and \ref{tab:RXTEcontinuum}). 

The possible change in the spectral description in CVs (from optically thin thermal plasma to up-scattering 
of soft photons due to inverse Compton in a thick Comptonization cloud) is analogous to adopting the Comptonization model to describe LMXB spectra in the $\sim$  3$-$50 keV energy range  
-- before Comptonization being broadly accepted, bremsstrahlung was the model used to describe the spectra 
of LMXBs \citep[see, e.g.,][]{Damico2001}. In a similar way, bremsstrahlung is the radiative process on the basis 
of the \textsc{mekal} and the cooling flow models which have been used to fit the spectral continuum of CVs. 
 In LMXBs the use of the Comptonization model  were then extended to broader (from $\sim$ 0.3 keV 
 to $\sim$ 250 keV) X-ray energy band \citep[see, e.g.,][]{DiSalvo2006,Dai2007, M09, Maiolino2013,Titarchuk2014}. 
 
 As we demonstrated in Fig. \ref{fig:plotGammakTeRXTE} 
 the  observed photon indices, $\Gamma$ is distributed around 1.8  using  the  data for  at least 15 spectra of nmCVs. Moreover,  in the previous section \ref{sec:model} we theoretically estimate the photon index of the emergent spectrum, $\Gamma$ which is formed in the TL around a WD.
 In order to do it we apply the radiative transfer formalism to solve 
 the boundary problem for the energy density distribution in the TL around a WD (see Eqs.  \ref{energy_distribution_ in_TL}$-$\ref{mod_epsilon_tau_1})  and estimate the gravitational energy release, $Q$ in the TL  in order to find   a Comptonization parameter presented by formula (\ref{y_parameter_estimate2}). Moreover, solving the eigenvalue problem for an average intensity $J(\tau)$ 
 (see Eq. \ref{eigenval_eq}) and using  a formula for the spectral index $\alpha$ ($\Gamma=\alpha+1$),  Eq. (\ref{index_alpha}) along with an estimate (\ref{y_parameter_estimate2}) we obtain that the photon index $\Gamma$ in the TL of an accreting  WD should be around 
 $1.85$. This is a first principle estimate of the photon index which is really confirmed by our analysis of the  nmCVs observations (see Fig. \ref{fig:plotGammakTeRXTE}).
 
  \subsection{Emission lines and continuum} 
The total observed spectrum consists of the continuum and line photons emission. In the standard 
framework, since the total spectrum shows many lines coming from different elements and ionisation 
degree, optically thin plasma with different temperatures, besides the one driving the continuum 
emission, are needed to describe all the spectral features. The different plasma temperature regions are located outside of the TL where the continuum is produced. As previously mentioned, spectral fittings using a single optically thin plasma 
are also used 
to describe the total spectrum. Though it can lead to a satisfactory fit, in terms of $\chi^2$ statistics, 
it can also lead to plasma temperatures too high to explain the presence of all emission lines. 
For example, for U~Gem a best-fit plasma temperature of about 16.5 keV 
(see section \ref{sec:UGem} and \citet{Xu2016}) is reported, at this temperature the plasma 
would be  completely ionized and no line(s) in the $\sim$ 6.4-6.7 keV range would be present. That is, at this temperature the $\sim$ 6.4-6.7 keV iron emission complex should not be produced in the same region of the X-ray continuum. 
Namely, much higher plasma temperatures are reported in IPs \citep{Xu2016}.  

In the framework of the continuum production through the thermal Comptonization, the continuum is formed in the TL -- in an optically thick medium, in both the quiescence and outburst states. 
The  emission lines are likely produced in an optically thin region, located in an external region, 
far from the TL and the WD surface. In other words, the continuum is formed  
closer to the WD than the lines. 

 Optically thin thermal plasma and cooling flow codes make simultaneously fits of the 
 continuum and the emission lines from several elements using  a range of temperatures (or few 
 components of optically thin thermal plasma of different temperatures). They  satisfactory  fit all (or almost all) excesses (features) superposing the continuum. 
For example, the \textsc{Mekal} model includes $\sim$2409 lines from all the 15 most 
important chemical elements (H, He, C, N, O, Ne, Na, Mg, Al, Si, S, Ar, Ca, Fe and Ni) 
and broad residual excesses might be fitted by several narrower emission lines. 
Changing  a description of the continuum to the Comptonization one  has a consequence 
that the residual excesses present in the spectra (i.e., the emission lines and their shape) 
will be directly observed after fitting the continuum -- as it is observed in LMXBs. 
In this case each of emission line should be independently identified and modeled using 
e.g., Gaussian components. 

Applying a thermal Comptonization component (either \textsc{compTT} or \textsc{compTB}) 
to fit the continuum of the nmCVs, we have identified  a broad and strong emission line 
peaked at  $\sim$ 0.96$-$1.02 keV in all \textit{XMM-Newton} spectra. 
We fitted this residual excess present in the $\sim$ 0.8 to 1.2 keV energy range with a single  
Gaussian line. We identified the centroid energy of the lines as being compatible with 
resonance lines emitted by Fe XXI (in  SS~Aur, U~Gem and VW Hyi), and/or Fe XVII 
(in VW Hyi, SS Aur and SS Cyg), and/or Ne X line (in VW Hyi and U Gem) 
(see Table \ref{tab:1lines}). This residual excess was also observed in one \textit{Chandra} 
HETG/ACIS spectrum of U~Gem in outburst (Obs. ID 3767), and in two SS~Cyg spectra in outburst. 
In these cases, the excess appears peaked at  $\sim 1.01 \pm 0.01$ keV. 
In U~Gem, the centroid energy of the Gaussian component fitted to this excess is compatible 
with Ne X, Fe XVII and Fe XXI lines. In SS Cyg, this excess is compatible with 
emission line from either Ca XVIII, Fe XVII, Ni XIX, or Ni XVIII in Obs. ID 648, {and Ca XVIII, Ni XIX, and Ne IX} in Obs. ID 2307. Other lines, compatible with emission from Mg XI, Mg XII, {Si XIII, and Fe XXII} 
 were also observed in the {\it Chandra} spectra, see Table \ref{tab:ChandraLines}. The 
 observation of this broad feature {around 1.01 keV} in both \textit{XMM-Newton} Epic-pn and  
 HETG/ACIS spectra, and its variability, indicates that this feature is not a systematic 
 effect of our spectral modeling, nor an individual issue of one of the instruments.   

It is important to point out that this broad excess likely correspond to a blend of 
narrow lines emitted by Fe XVII, XIX, XX, and XXI and/or Ne IX-X, and/or Ni XIX ions -- which 
are not very well resolved by the medium energy resolution of the pn camera and HETG/ACIS. 
Several lines are observed  in this soft X-ray energy range by LETG/HRC grating spectrometer 
(due its resolving power $>$ 1000) \cite[see also, e.g.][]{Mukai2003,Pandel2003}. For example, 
in our analysis of the LETG/HRC SS Cyg spectrum, lines at $\sim$ 0.82 keV, 
$\sim 0.92$ keV, and $\sim$ 1.0 keV were clearly present.

\section{Summary and Conclusions}
\label{sec:conclusions}

The thermal Comptonization model plus Gaussian components (used to account for the emission lines) can successfully describe the spectra of the nmCVs in our source sample. 
{Though we did not find a perfect fit, in terms of $\chi^2$ statistics, to  
the \textit{XMM-Newton} spectra of VW~Hyi and U~Gem, both the \textit{Chandra} 
HETG/ACIS and \textit{RXTE} PCA spectra of the later source were satisfactory fitted by our Comptonization model. It is likely that the presence of many lines, besides the iron complex, in the XMM-Newton spectra of VW~Hyi and U~Gem worsened the fit quality.
}

The \textit{XMM-Newton} and \textit{Chandra} spectra show a similar 
range of physical parameters. 
However, the \textit{RXTE} spectra, due to the broader spectral energy range, 
provided a better description  of the Comptonization effect 
and determination of the physical parameters.
We found  that two Comptonization components are necessary to successfully fit
the simultaneous  0.4$-$150 keV \textit{Chandra}/\textit{RXTE} spectra of SS~Cyg 
in quiescence, and two 2.5$-$25 keV \textit{RXTE} spectra of SS~Cyg in outburst. 
In this case, the best-fits are found only 
if the optical depth and the plasma temperature of the Compton cloud are  the same for these two components. As a results  we found  the only one photon index $\Gamma\sim1.8$   is capable to describe the total spectra  of all analyzed nmCVs. 

Two blackbody components of seed photons  
are characterized by their color temperature (kT$_{s1}$ and kT$_{s2}$). 
In our interpretation, the seed photon component showing lowest temperatures (kT$_{s1} \sim$ 0.1$-$0.2 keV) 
is presumably coming from the inner part of the accretion disk {or outer part of the TL}, 
while the second and transient component related to $kT_{s2}$ 
is coming from a more internal part of the system, closer to the WD surface.  We rather think that these temperatures are related to  the innermost part of the corona (TL)  than they are associated with a WD surface, otherwise a value of kT$_{s2} \sim$ 
0.6$-$0.8 keV {could} exceeds the appropriate temperature limit for the existence of bounded atmospheres on WDs. 

The seed photons are up scattered by hot electrons in the TL 
characterized by a single electron temperature $kT_e$ (in the 5$-$48 keV range), 
and located between the disk and the WD surface. 
The TL optical depth changes in the wide range $1\lax\tau_0\lax5$.

When we compared the physical parameters between different sources in different states, we did not find any correlation of $\Gamma$
(or $\alpha=\Gamma-1$) with the plasma temperature $kT_e$ (see Fig \ref{fig:plotGammakTeRXTE}), nor a clear correlation of physical parameters with the source stage. However, the \textit{RXTE} observations of SS~Cyg 
show a clear change of the electron temperature in the Compton cloud (or TL) 
when the source is found in different states: it is higher in quiescence, 
33$^{+3}_{-3}$ keV;  it decreases to 25$^{+1}_{-1}$ keV at the initial stage 
of the optical outburst;  and reaches $\sim 5.5^{+0.4}_{-0.3}$ keV during a rise of the 
outburst. This decrease in the electron temperature can explain the suppression 
of hard X-rays ($\gtrsim$ 25 keV) during outburst in SS Cyg. Therefore,  the dependence of the physical parameters with the source stages may depend on the source. 

Finally, we conclude that our two thermal Comptonization component model  using  a  single thermal plasma temperature
and optical depth  plus Gaussian components 
can  describe the 0.4$-$150 keV spectra of nmCVs, in both quiescence and outburst states, without evoking  more than one optically thin plasma temperature or cooling flow models. 

Moreover, we develop  the radiative transfer model which rigorously demonstrates and explains our main observational result, using the  first principal arguments, that the photon indices, $\Gamma$ in WDs should be around 1.8 (see Fig. \ref{fig:plotGammakTeRXTE}).

\section*{Acknowledgments}
{ 
T.M.\ acknowledges the financial support given by the Erasmus Mundus 
Joint Doctorate Program by Grants Number 2013-1471 from the agency 
EACEA of the European Commission, and the INAF/OAS Bologna. 
T.M.\ acknowledges the support given by the National Program on 
Key Research and Development Project (Grants No. 2016YFA0400803) 
and the NSFC (11622326 and U1838103). T.M.\ also thanks the High Energy Astrophysics group of the Physics Dept.\ of the University of Ferrara, INAF/OAS Bologna and Wuhan University for the warm hospitality and support.  L.T.\ appreciates the interest and support of his colleagues from the Physical Institute of the Russian Academy of Science (FIAN). M.O.\ acknowledges support from the Italian Space Agency under grant ASI-INAF 2017-14-H.0. The authors are grateful to the anonymous referee for the constructive suggestions.
}

\bibliographystyle{apj}
\bibliography{biblio}
\end{document}